\documentclass[aps, prd, superscriptaddress, nofootinbib, preprintnumbers]{revtex4}
\usepackage{amsmath}
\usepackage{graphicx}

\newcommand{\lesssim}{\mathrel{\mathpalette\vereq<}}
\newcommand{\gtrsim}{\mathrel{\mathpalette\vereq>}}

\newcommand{\chushi}[1]{}
\newcommand{\eqn}[1]{&\hspace{-0.6em}#1\hspace{-0.6em}&}

\begin{document}
\begin{flushright}
SU-HET-02-2016%
\end{flushright}

\title{Invisible Axion-Like Dark Matter from Electroweak Bosonic Seesaw}       
\author{Hiroyuki Ishida}\thanks{{\tt ishida@riko.shimane-u.ac.jp}}
	\affiliation{Graduate School of Science and Engineering, Shimane University,
 Matsue 690-8504, Japan.}
\author{Shinya Matsuzaki}\thanks{{\tt synya@hken.phys.nagoya-u.ac.jp}}
      \affiliation{ Institute for Advanced Research, Nagoya University, Nagoya 464-8602, Japan.}
      \affiliation{ Department of Physics, Nagoya University, Nagoya 464-8602, Japan.}   
\author{Yuya Yamaguchi}\thanks{{\tt yy@particle.sci.hokudai.ac.jp}}
	\affiliation{Graduate School of Science and Engineering, Shimane University,
 Matsue 690-8504, Japan.}
	\affiliation{Department of Physics, Faculty of Science, Hokkaido University,
 Sapporo 060-0810, Japan.}


\begin{abstract}
We explore a model based on
the classically-scale invariant standard model (SM) with a strongly coupled vector-like dynamics,
which is called hypercolor (HC).
The scale symmetry is dynamically broken by the vector-like condensation at the TeV scale,
so that the SM Higgs acquires the negative mass-squared by the bosonic seesaw mechanism
to realize the electroweak symmetry breaking.
An elementary pseudoscalar $S$ is introduced
to give masses for the composite Nambu-Goldstone bosons (HC pions):
the HC pion can be a good target to explore through a diphoton channel at the LHC.
As the consequence of the bosonic seesaw, the fluctuating mode of $S$, which we call $s$,
develops tiny couplings to the SM particles and is predicted to be very light. 
The $s$ predominantly decays to diphoton
and can behave as an invisible axion-like dark matter.
The mass of the $s$-dark matter is constrained by currently available cosmological and astrophysical limits 
to be $10^{-4} {\rm eV} \lesssim m_s \lesssim 1 \,{\rm eV}$.
We find that the sufficient amount of relic abundance for the $s$-dark matter  
can be accumulated via the coherent oscillation. 
The detection potential in microwave cavity experiments is also addressed.

\end{abstract} 
\maketitle

\section{Introduction}

Discovery of the Higgs boson~\cite{Aad:2012tfa,Chatrchyan:2013lba}  
has made up for the last piece of the standard model (SM) in terms of the particle content, 
though the Higgs physics such as the coupling property is still uncertain. 
The Higgs boson in the SM plays the important role for the electroweak symmetry breaking, which 
is triggered by the nonzero vacuum expectation value of the Higgs field with the negative mass-squared.      
The Higgs mass term can be necessarily introduced once the Higgs field is put in, due to the renormalizability 
upon which the SM has been established. 
However, the Higgs mass term still involves unsatisfactory ingredients on the theoretical ground:    
one is called the gauge hierarchy problem, that is what we need to answer: how  to stabilize the electroweak vacuum.  
The other, which would be correlated with the former, is the origin of the ``negative''-mass squared, 
which is simply assumed in the SM without any dynamical concept.

One intriguing idea to solve those problems 
is to assume 
the classical-scale invariance in the SM. 
In this approach, there is no dimensionful parameter at a classical level, 
so one does not need to take care of quadratic divergent terms regarding 
the renormalization of the Higgs boson mass: thus no scale is present in the model, which would be anomalously generated, 
e.g., via the radiative breaking. 
The classical-scale symmetry is then broken by the Coleman-Weinberg mechanism~\cite{Coleman:1973jx},  
which generates the mass scale via dimensional transmutation.

Actually, such a radiative-breaking scenario does not work solely within the SM itself, due to the presence of the heavy top quark, 
so one is eventually forced to add extra degrees of freedom to trigger the radiative breaking as desired.    
One of the way out along this approach would be to extend the SM gauge symmetry by introducing an extra $U(1)$ gauge symmetry~\cite{Hempfling:1996ht}.  
However, most of such models suffer from an ad hoc assumption: 
one requires to assume the sign of the quartic coupling between the SM Higgs boson 
and an additional scalar field to be negative.

This ``sign problem" can be solved in a way of a dynamical mechanism, which is called bosonic seesaw mechanism~\cite{BSS}. 
The mechanism itself is essentially analogue to the usual type-I seesaw mechanism, well-known in addressing the neutrino sector. 
The key point to note is that in the case of fermions, the phase of mass term can be absorbed by redefinition of the fermion fields, 
just by the $U(1)_A$ phase rotation.  
In contrast, the phase cannot be removed by any ways for boson mass terms, rather will be physical 
once the negative sign shows up in the bosonic sector.  
Thus models having the bosonic seesaw mechanism built based on the classical-scale invariance 
can realize the desired situation in which the problems raised above can be settled down by a dynamical explanation.

Recently, such a hybrid model encoding both the classical-scale invariance and bosonic seesaw mechanism  
has been proposed~\cite{Haba:2015qbz}. 
The model in Ref.~\cite{Haba:2015qbz} is constructed based on  
the classically-scale invariant SM plus a vector-like strongly coupled sector, which we call hypercolor (HC) 
(which was originally quoted as ``technicolor" in Ref.~\cite{Haba:2015qbz}).   
In the model the classical-scale invariance is dynamically broken by the vector-like condensation 
of the HC fermion bilinear, triggered by the strongly coupled HC, in a way analogous to QCD. 
The negative mass-squared of the Higgs is induced by the bosonic seesaw mechanism through 
the mixing between the elementary Higgs doublet and the composite-HC Higgs doublet 
formed as the HC fermion bound state. 
Thus, the HC dynamics plays the essential role to solve both the gauge hierarchy and negative-mass squared problems.

In the model of~\cite{Haba:2015qbz}   
the success of the bosonic seesaw by the HC dynamics is subject to the presence of the ``chiral" symmetry 
carried by the HC fermions, which is partially vector-like gauged by the electroweak charges to provide 
the composite-HC Higgs doublet with the appropriate SM charges.    
When the HC fermion condensate develops to be nonzero, the ``chiral" symmetry is dynamically broken 
down to the vectorial subgroup, at the same time the scale symmetry is broken. 
This leads to a couple of Nambu-Goldstone bosons (HC pions).

The origin of mass for the HC pions in the model of~\cite{Haba:2015qbz}  
is responsible for a pseudoscalar $S$, having the Yukawa coupling to the HC fermions explicitly 
breaking the ``chiral" symmetry: the pseudoscalar $S$ develops the nonzero vacuum expectation value as the direct 
consequence of the bosonic seesaw, to give the HC pion masses via the Yukawa coupling, 
hence acts as another ``Higgs" for the HC pions. 
Thus, discovering the pseudoscalar $S$ as well as the HC pions is a direct probe  
and a smoking-gun for this bosonic seesaw model.

In this paper, we discuss the phenomenological consequence of the pseudoscalar $S$ 
linking to the presence of the HC pions, crucial for the bosonic seesaw model of Ref.~\cite{Haba:2015qbz}. 
We find that, after the dynamical-scale breaking and triggering the electroweak bosonic seesaw 
at a TeV scale, the fluctuating mode of $S$, which is denoted as $s$, 
develops vanishingly small couplings to the SM particles.  
It turns out, furthermore, that in close relation to the HC pion masses 
the $s$ mass is predicted to be very light and to predominantly decay to diphoton. 
We then identify the $s$ as a dark matter candidate like an invisible axion-like particle  
and constrain the $s$ mass by several cosmological and astrophysical bounds. 
The $s$ mass is thus limited to be $10^{-4} \, {\rm eV} \lesssim m_s \lesssim 1 \, {\rm eV}$.

We examine the possibility of the cosmological productions of the $s$  
and show that the $s$ is unlikely to be thermally produced essentially 
due to its tiny couplings to the HC sector in the thermal equilibrium.  
We then find that the sufficient amount of relic abundance of $s$ as the cold dark matter 
can be accumulated via the coherent oscillation, just like the invisible axion case. 
The detection potential in microwave cavity experiments is also addressed.  
It is shown that the $s$ with mass around $1 \, {\rm eV}$ can have the same level of the 
detection sensibility as that of the axion in the currently equipped experimental 
setup, so the $s$ is detectable by the microwave cavity experiments.

This paper is organized as follows: 
in Sec.~\ref{HC-model} we first review the model of Ref.~\cite{Haba:2015qbz} 
by focusing on the essential points to realize the electroweak symmetry breaking 
via the bosonic seesaw and to give masses to the HC pions. 
(The details for the calculation of the HC pions signals are given in a couple of Appendices.) 
In Sec.~\ref{s-DM} we show the close relationship between masses of 
the pseudoscalar $s$ and the HC pions, and the $s$ couplings to the SM particles 
arising as the direct consequence of the bosonic seesaw mechanism, which turns out to be vanishingly small. 
We then identify the $s$ as a dark matter candidate and constrain the mass 
by several cosmological and astrophysical bounds currently at hand. 
In Sec.~\ref{prod-s} we discuss the cosmological productions of the $s$ involving  
the thermal and non-thermal processes. 
It is shown that the relic abundance of the $s$ cannot be thermally produced enough to account for 
the present dark matter density due to the tiny couplings to the SM particles. 
We then find that the non-thermal production, namely, the coherent oscillation 
is dominant in the production mechanism for the $s$, which is sufficient for 
the $s$ to be a cold dark matter in the present universe. 
The detection potential of the $s$-dark mater in microwave cavity experiments 
are also discussed in comparison with the case of invisible axion-like particles. 
Summary and discussion are given in Sec.~\ref{summary}.

Appendix~\ref{Pi-mass} provides the details for computation of the HC pion masses, 
and Appendix~\ref{chiral-lag} gives derivation of the HC pion couplings 
based on the nonlinear realization of the ``chiral" symmetry. 
The $s$ couplings are also generated there due to the mixing with the HC eta-prime 
arising through the bosonic seesaw. 
In Appendix~\ref{750} we present the decay properties of the HC pions 
relevant to the LHC study, and show the details of the LHC production cross sections 
to compute the 750 GeV HC pion signals, in comparison with the current LHC bounds. 

\section{A hypercolor model with bosonic seesaw mechanism}
\label{HC-model}

The model we employ is based on the classically-scale invariant SM plus 
a strongly coupled HC dynamics at the TeV scale.   
The way to construct the model follows from the literature~\cite{Haba:2015qbz}. 
The HC sector is described by the HC-gluon ${\cal G}$ with a gauge coupling $g_{\rm HC}$ and 
three vector-like fermion triplets ,  
$ F_{L,R} = (\chi_i, \psi)^T_{L,R}$, 
having the charges, 
$ \chi_{i(i=1,2)} \sim (N_{\rm HC}, 1, 2, 1/2)$
and $\psi \sim (N_{\rm HC}, 1, 1, 0)$
for the HC group $SU(N_{\rm HC})$ and $SU(3)_c \times SU(2)_W \times U(1)_Y$. 
The HC theory possesses the ``chiral" $U(3)_L \times U(3)_R$ symmetry as well as the (classically) scale-invariance. 
The main part of the model Lagrangian thus goes like 
\begin{equation} 
 {\cal L} 
 = 
 {\cal L}_{\rm SM}|_{m_H=0} 
 + 
 \bar{F} i \gamma^\mu D_\mu F - \frac{1}{2} {\rm tr}[{\cal G}_{\mu\nu}^2] 
 - V
 \,, \label{L}
\end{equation}
with 
\begin{equation} 
 D_\mu = \partial_\mu - i g_{\rm HC} {\cal G}_\mu 
\,, \qquad 
{\cal G}_{\mu\nu} 
= \partial_\mu {\cal G}_\nu - \partial_\nu {\cal G}_\mu 
- ig_{\rm HC} [{\cal G}_\mu , {\cal G}_\nu] 
\,.  
\end{equation}
Here the SM gauges have been switched off momentarily and the potential term $V$ 
will be specified later.

The ``chiral" symmetry is assumed to be explicitly broken due to the 
the breaking terms: 
\begin{eqnarray} 
\Delta {\cal L}' &=& 
{\cal L}_y + {\cal L}_S 
\,, \\ 
{\cal L}_y  
&=& 
-  
y 
\, 
\bar{F}_L \cdot   
\left( 
\begin{array}{cc} 
0 & H \\ 
H^\dag & 0   
\end{array} 
\right) 
\cdot 
F_R 
+ {\rm h.c.} 
\,,   \label{y}\\ 
{\cal L}_S 
&=& 
 i g_S 
\left( 
\bar{F}_L F_R - \bar{F}_R F_L   \right)S 
\,,\label{gS-term} 
\end{eqnarray} 
where the Yukawa and $g_S$ couplings $y$ and $g_S$ are assumed to be $\ll 1$  
in order to realize the ``chiral" symmetry approximately;   
$H$ denotes the elementary Higgs doublet, and the $S$ is a pseudoscalar field having no SM charges.  
The potential term $V$ in Eq.(\ref{L}) includes the $H$ and $S$ like 
\begin{equation} 
 V = \lambda_H (H^\dag H)^2 
+ \kappa_H S^2 (H^\dag H) 
+ \lambda_S S^4 
\,. \label{V} 
\end{equation}  
Thus, the full Lagrangian terms are constructed from Eqs.(\ref{L}), (\ref{y}), (\ref{gS-term}) and (\ref{V}) 
as 
$ {\cal L} + \Delta {\cal L}' $.

Among the ``chiral'' symmetry, 
$U(1)_A$ is to be explicitly broken by the anomaly, and the remaining  (approximate) ``chiral" 
$SU(3)_L \times SU(3)_R (\times U(1)_{V})$ is broken by the ``chiral'' condensate, invariant 
under the SM gauge symmetry,  
$\langle \bar{F}F  \rangle = \langle \bar{\chi}_i \chi_i \rangle = \langle \bar{\psi} \psi \rangle \neq 0$, 
down to the diagonal subgroup 
$SU(3)_{V} (\times U(1)_{V})$ at the strong scale $\Lambda_{\rm HC}$, just like 
the ordinary QCD.  
The ``chiral" condensate $\langle \bar{F}F \rangle$  
then gives rise to the 8 Nambu-Goldstone bosons (plus heavy $\eta'$).

\subsection{Scalar Seesaw}

 At the $\Lambda_{\rm HC}$ scale the composite HC Higgs fields $\sim \bar{F}_iF_j$ 
are generated. Among them, the component $\Theta \sim \chi \bar{\psi}$ has the 
same quantum number as that of the elementary Higgs doublet $H$. 
The mixing between the $\Theta$ and $H$ thus gives rise to the scalar seesaw~\cite{Haba:2015qbz}.

Taking into account 
the Yukawa term ${\cal L}_y$ in Eq.(\ref{y}) and generation of 
the $\Theta$ mass term,   
one can write the effective Lagrangian at $\Lambda_{\rm HC}$ to quadratic order in  
fields 
as 
\begin{equation} 
{\cal L}_{\rm eff}(\Lambda_{\rm HC})  
= - y \left[ \Theta^\dag \cdot H + {\rm h.c.} \right] 
- M_{\Theta}^2 \Theta^\dag \Theta 
\,. \label{eff:Lag}
\end{equation} 
This leads to the seesaw type mass matrix for the Higgs doublet $H$ and 
the composite Higgs doublet $\Theta$ (``bosonic seesaw"): 
\begin{equation}  
\left( 
\begin{array}{c} 
H \\ 
\Theta
\end{array}
\right)^\dag 
 \left( 
\begin{array}{cc} 
0 & y \Lambda^2_{\rm HC} \\ 
y \Lambda^2_{\rm HC} & M_{\Theta}^2 
\end{array}
\right)  
\left( 
\begin{array}{c} 
H \\ 
\Theta
\end{array}
\right)
\,. 
\end{equation}
This is diagonalized by expanding terms in powers of $y \ll 1$ 
to be 
\begin{equation} 
\left( 
\begin{array}{c} 
H_1 \\ 
H_2
\end{array}
\right)^\dag 
 \left( 
\begin{array}{cc} 
- y^2 \frac{\Lambda_{\rm HC}^4}{M_\Theta^2}  & 0  \\ 
0 & M_\Theta^2 (1 + \frac{y^2 \Lambda_{\rm HC}^2}{M_\Theta^2})    
\end{array}
\right)  
\left( 
\begin{array}{c} 
H_1 \\ 
H_2
\end{array}
\right)
\equiv 
\left( 
\begin{array}{c} 
H_1 \\ 
H_2
\end{array}
\right)^\dag 
 \left( 
\begin{array}{cc} 
- m_{H_1}^2 & 0  \\ 
0 & m_{H_2}^2   
\end{array}
\right)  
\left( 
\begin{array}{c} 
H_1 \\ 
H_2
\end{array}
\right)
\,. 
\end{equation}
The mass eigenstates $(H_1, H_2)$ are related to the current eigenstates $(H, \Theta)$ as  
\begin{equation} 
\left( 
\begin{array}{c} 
H_1 \\ 
H_2 
\end{array}
\right) 
\simeq 
\left(
\begin{array}{cc} 
1 - \frac{y^2}{2} + {\cal O}(y^4) & - y (1 - \frac{3}{2} y^2) + {\cal O}(y^5) \\ 
y (1 - \frac{3}{2} y^2) + {\cal O}(y^5)& 1 - \frac{y^2}{2} + {\cal O}(y^4)   
\end{array}
\right)
\left( 
\begin{array}{c} 
H \\ 
\Theta 
\end{array} 
\right)
\,, \label{relation:H}
\end{equation} 
where we have taken $M_{\Theta} \simeq \Lambda_{\rm HC}$. 
Thus the scale-breaking effect has been transfered to the $H$-Higgs sector via 
the bosonic seesaw mechanism. 
Note the negative sign for the lower eigenvalue $(- m_{H_1}^2)$,  
playing the essential role to realize the electroweak symmetry breaking,   
as will be explicitly clarified later on.

\subsection{Pseudoscalar Seesaw}

As mentioned above, the $\eta'$ gets the mass from the $U(1)_A$ anomaly as in the case of 
the ordinary QCD. 
The size of the mass 
can be estimated just by scaling 
from the QCD to be 
\begin{equation} 
M_{\eta'} 
\sim {\cal O}(1{\rm GeV}) \times \left( \frac{\Lambda_{\rm HC}}{\Lambda_{\rm QCD}} \right) \times 
\sqrt{ \frac{3}{N_{\rm HC}}} 
\sim {\cal O}(1\, {\rm TeV} ) \times \sqrt{\frac{3}{N_{\rm HC}}} 
\,, 
\end{equation} 
where the large $N_{\rm HC}$ counting has been taken into account.  
One should note that 
the $\eta'$ couples to the $U(1)_A$ current, 
$J^0_\mu = \frac{1}{\sqrt{6}}\cdot \bar{F} \gamma_\mu \gamma_5 \cdot {\bf 1}_{3 \times 3} \cdot F$.  
Hence at the $\Lambda_{\rm HC}$ scale, by taking into account the $\eta'$ mass generation from the anomaly,  
the $g_S$ term  in Eq.(\ref{gS-term}) looks  like 
\begin{equation} 
 {\cal L}_S(\Lambda_{\rm HC}) 
\approx  
g_S \Lambda^2_{\rm HC} \eta' S  - \frac{1}{2}M^2_{\eta'} (\eta')^2  
\,. \label{eff:Lag:eta}
\end{equation}
Again, the form of Eq.(\ref{eff:Lag:eta}) is nothing but a seesaw type (``bosonic seesaw"), 
so one can readily see that the lower eigenvalue, corresponding to the $S$-mass squared, 
is negative: 
\begin{eqnarray} 
 {\cal L}_S(\Lambda_{\rm HC}) 
&\approx& 
- \frac{1}{2}
\left( 
\begin{array}{c} 
S \\ 
\eta'
\end{array}
\right)^T 
\left( 
\begin{array}{cc} 
 0 & - g_S \Lambda_{\rm HC}^2 
\\ 
- g_S \Lambda_{\rm HC}^2 & M_{\eta'}^2  
\end{array}
\right)     
\left( 
\begin{array}{c} 
S \\ 
\eta'
\end{array}
\right) 
\nonumber \\ 
&=& 
- \frac{1}{2}   
\left( 
\begin{array}{c} 
{\cal S} \\ 
\eta^0
\end{array}
\right)^T  
 \left( 
\begin{array}{cc} 
- g_S^2 \frac{\Lambda_{\rm HC}^4}{M_{\eta'}^2} & 0  \\ 
0 & M_{\eta'}^2 (1 + \frac{g_S^2 \Lambda_{\rm HC}^2}{M_{\eta'}^2})   
\end{array}
\right)  
\left( 
\begin{array}{c} 
{\cal S} \\ 
\eta^0 
\end{array}
\right)
\equiv 
\left( 
\begin{array}{c} 
{\cal S} \\ 
\eta^0
\end{array}
\right)^T  
 \left( 
\begin{array}{cc} 
- m_{\cal S}^2 & 0  \\ 
0 & m_{\eta^0}^2   
\end{array}
\right)  
\left( 
\begin{array}{c} 
{\cal S} \\ 
\eta^0 
\end{array}
\right)
\,. 
\end{eqnarray}
The mass eigenstates $({\cal S}, \eta^0)$ are related to the current eigenstates $(S, \eta')$ as  
\begin{equation} 
\left( 
\begin{array}{c} 
{\cal S} \\ 
\eta^0
\end{array}
\right) 
\simeq 
\left(
\begin{array}{cc} 
1 - \frac{g_S^2}{2} + {\cal O}(g_S^4) & g_S (1 - \frac{3}{2} g_S^2) + {\cal O}(g_S^5) \\ 
- g_S (1 - \frac{3}{2} g_S^2) + {\cal O}(g_S^5) & 1 - \frac{g_S^2}{2} + {\cal O}(g_S^4)   
\end{array}
\right)
\left( 
\begin{array}{c} 
S \\ 
\eta' 
\end{array} 
\right)
\,, \label{relation:S}
\end{equation} 
to the nontrivial order of expansion in $g_S \ll 1$, 
where we have taken $M_{\eta'} \simeq \Lambda_{\rm HC}$. 
Thus, the pseudoscalar ${\cal S}$ can get the nonzero vacuum expectation value, 
playing the significant role to supply the pseudo Nambu-Goldstone boson (HC pion)  masses, 
as will be clearly seen later.

\subsection{Electroweak Symmetry Breaking}

Including the dynamically generated terms, 
we thus see that 
Eq.(\ref{V}) is now modified 
at the scale $\Lambda_{\rm HC}$ 
as follows: 
\begin{eqnarray} 
V 
&=&  
- 
\left( 
\begin{array}{c} 
H \\ 
\Theta
\end{array}
\right)^\dag 
 \left( 
\begin{array}{cc} 
0 & y \Lambda^2_{\rm HC} \\ 
y \Lambda^2_{\rm HC} & M_{\Theta}^2 
\end{array}
\right)  
\left( 
\begin{array}{c} 
H \\ 
\Theta
\end{array}
\right)  
- \frac{1}{2}
\left( 
\begin{array}{c} 
S \\ 
\eta'
\end{array}
\right)^T 
\left( 
\begin{array}{cc} 
 0 & g_S \Lambda_{\rm HC}^2 
\\ 
g_S \Lambda_{\rm HC}^2 & M_{\eta'}^2  
\end{array}
\right)     
\left( 
\begin{array}{c} 
S \\ 
\eta'
\end{array}
\right) 
\nonumber \\ 
&& 
+ \lambda_{\Theta} (\Theta^\dag \Theta)^2 
+ \lambda_{H} (H^\dag H)^2 
+ \kappa_H S^2 (H^\dag H) 
+ \lambda_S S^4 
 \nonumber \\ 
&=& 
-  m_{H_1}^2 (H_1^\dag H_1) + m_{H_2}^2 (H_2^\dag H_2) 
- \frac{1}{2} m_{\cal S}^2 {\cal S}^2 + \frac{1}{2} m_{\eta^0}^2 (\eta^0)^2
\nonumber \\ 
&& 
+ \lambda_{\Theta} (\Theta^\dag \Theta)^2 
+ \lambda_{H} (H^\dag H)^2 
+ \kappa_H S^2 (H^\dag H) 
+ \lambda_S S^4 
\,, 
\label{pot}
\end{eqnarray} 
where we added the quartic coupling of $\Theta$ which can generically be induced from 
the underlying HC dynamics, and is expected to be $\gtrsim {\cal O}(10)$.  
 Based on this potential we discuss the realization of the electroweak symmetry breaking.

To this end, 
we may first parametrize the scalar and pseudoscalar fields with their vacuum expectation values 
for the mass eigenstate fields $(H_1, H_2)$ and $({\cal S}, \eta^0)$ in Eqs.(\ref{relation:H}) 
and (\ref{relation:S}):  
\begin{eqnarray} 
H_1 
&=& 
\left(
\begin{array}{c}
 \varphi_1^+ \\ 
 \frac{1}{\sqrt{2}} (v_1 + h_1^0 + i \varphi_1^0)  
\end{array}
\right) 
\,, \qquad 
H_2=
\left(
\begin{array}{c}
 \varphi_2^+ \\ 
 \frac{1}{\sqrt{2}} (v_2 + h_2^0 + i \varphi_2^0)  
\end{array}
\right) 
\,, \nonumber \\ 
{\cal S} 
&=& v_S + s 
\,, \qquad 
\eta^0 = v_{\eta} + e_0 
\,, \label{field:para}
\end{eqnarray} 
where $\pm$ denote the electromagnetic charges assigned according to the charges of the HC-$F$ fermions. 
We may search for the vacuum by assuming\footnote{The stationary condition for $v_2$ actually includes the trivial solution $v_2 =0$,  hence one can always select the vacuum with $v_2=0$ which in the present study we have taken for simplicity. 
Under the condition with $v_2=0$, however, other vacuum expectation values ($v_S$,$v_\eta$) cannot be set to zero because of some phenomenological constraints, related to the HC pion and eta-prime masses, as will be seen later (See Eqs.~(\ref{stationary:condi}), (\ref{masses}) and (\ref{gsvs})).} 
\begin{equation}
 v_2 = 0 
\,, \label{v2:0}
\end{equation}
so that, for the nontrivial solutions $v_1\neq 0, v_S \neq 0, v_\eta \neq 0$, 
 the stationary conditions are obtained by expanding terms in powers of $y$ and $g_S$ as  
\begin{eqnarray}
m_{H_1}^2 
&=& \frac{1}{2} y^2 \lambda_\Theta v_1^2 + \cdots  
(\simeq y^2 \Lambda_{\rm HC}^2) 
\,, \nonumber \\ 
m_{\cal S}^2 
&=& 4 \lambda_S v_S^2 + \cdots 
(\simeq g_S^2 \Lambda_{\rm HC}^2) 
\,, \nonumber \\ 
m_{\eta^0}^2 
&=& 
g_S^3 \frac{v_S^3}{v_\eta} + \cdots 
(\simeq \Lambda_{\rm HC}^2)
\,, \nonumber \\ 
\kappa_H &=& - \frac{v_1^2}{v_S^2} 
\lambda_H 
+ \cdots 
\,,  \label{stationary:condi}
\end{eqnarray}
where the last condition has come by imposing $v_2=0$ and 
the ellipses denote terms suppressed by higher orders in expansion with respect to 
$y$ and $g_S$, and  
the expressions in the parenthesis correspond to the seesaw-induced formulae. 
As will be discussed in the later section, the $v_S$ is constrained, by the 
phenomenological limits on the pseudoscalar $s$, as $\Lambda_{\rm HC}/v_S \ll 1$, 
so that the coupling $\kappa_H$ is required to be vanishingly small, $\kappa_H \ll 1$, 
hence so is the $\lambda_S$, $\lambda_S \ll 1$.

By adjusting parameters to satisfy these conditions,  
the electroweak scale $v_1 = 246$ GeV can be realized at the minimum of the 
potential (with the $H$-quartic coupling $\lambda_H >0$, hence $\kappa_H <0$), 
consistently with the bosonic seesaw mechanism.

As will be clarified later (Eq.(\ref{masses})), 
the square of masses for fluctuating fields $(h_1^0, h_2^0, s, e_0)$ are properly 
positive-definite at the chosen stationary space $(v_1, v_2, v_S, v_\eta)$ 
satisfying the stationary conditions Eq.(\ref{stationary:condi}) 
with $v_2=0$. This implies that the vacuum has safely been aligned to where the electroweak symmetry is broken with extra nonzero 
CP-odd vacuum expectation values $(v_S, v_\eta)$.  
By taking some reference values for the potential parameters, 
we have numerically checked that the electroweak-broken vacuum 
indeed locates at the global minimum. 
Actually, the alignment problem should be 
argued by taking into account all the possible vacuums including 
nonzero vacuum expectation values for other composite HC Higgs fields like 
$\bar{\chi}\chi, \bar{\psi}\psi$, and so forth. 
However, due to the presence of the ``chiral'' symmetry in the underlying
HC theory, one can be allowed to rotate the composite HC Higgs fields to be 
aligned to the desired direction where the potential 
is minimized at the electroweak-broken vacuum. 
More rigorous proof is to be beyond scope of the present study, which 
will be argued elsewhere.

\subsection{Scalar and Pseudoscalar Masses} 

The scalars ($h_1, h_2$) and pseudoscalars ($s, e_0$), defined as in Eq.(\ref{field:para}), 
arise as the fluctuating modes around the vacuum expectation values $(v_1, v_S, v_\eta)$  
in the potential Eq.(\ref{pot}). 
Expanding the potential terms in powers of the small parameters $(y, g_S, v_1/v_S, \kappa_H, \lambda_S)$ and keeping only the nontrivial leading orders, one finds 
the mass eigenvalues,   
\begin{eqnarray} 
 m_{h_1^0}^2  &\simeq& 
 2 \lambda_H v_1^2 \simeq 2 (-\kappa_H) v_s^2
 \,, \nonumber \\ 
 m_{h_2^0}^2 &\simeq& m_{H_2}^2 
 \,, \nonumber \\ 
 m_{s}^2 &\simeq&  8 \lambda_S v_S^2 \simeq 2 g_S^2 \Lambda_{\rm HC}^2 
 \,, \nonumber \\ 
m_{e_0}^2 &\simeq & m_{\eta^0}^2 
\,, \label{masses}
\end{eqnarray}
where the second approximate expression in the third line  
follow from the stationary conditions in Eq.(\ref{stationary:condi}) and 
the $h_1^0$ is identified as the 125 GeV Higgs. 
It is interesting to note that, in addition to particles with the ${\cal O}({\rm TeV})$ mass on the natural scale of HC dynamics, 
the present model predicts a light pseudoscalar $(s)$ with mass of ${\cal O}(g_S \Lambda_{\rm HC}) (\ll \Lambda_{\rm HC})$, 
as the consequence of the bosonic seesaw mechanism. 
Thus, this $s$ is a smoking-gun of the model and will be identified as the dark matter 
candidate, as will be discussed later.

\subsection{HC pions}

Since the $y-$ and $g_S-$ Yukawa terms in Eqs.(\ref{y}) and (\ref{gS-term}) 
explicitly break the ``chiral'' $SU(3)_L \times SU(3)_R$ symmetry, 
the 8 Nambu-Goldstone bosons become pseudo's (HC pions $\Pi$) through those interactions. 
Using the current algebra technique and expanding things in powers of $y$ and $g_S$, 
 one can evaluate the HC pion masses to 
find that they are almost degenerate to be   
\begin{equation} 
 m_{\Pi} \simeq 2 (g_S v_S) \frac{\Lambda_{\rm HC}}{f}
\,, \label{HC-pion-mass}
\end{equation}
where 
\begin{equation} 
 f = \frac{f_\Pi}{\sqrt{N_{\rm HC}/3}}
\,,  
\end{equation}
with the $f_\Pi$ being the HC pion decay constant. 
The detail of the derivation for this formula is presented in Appendix~\ref{Pi-mass}.

As a reference point, we may set the HC pion mass to be $750$ GeV so that 
the combination $(g_S v_S)$ can be fixed as 
\begin{equation} 
 (g_S v_S) \simeq 
 30\, {\rm GeV} \times \left( \frac{m_{\Pi}}{750\,{\rm GeV}}  \right) 
\left( \frac{4 \pi f}{\Lambda_{\rm HC}}  \right) 
\,.  \label{gsvs}
\end{equation}
We may take $\Lambda_{\rm HC} \sim 4 \pi f$ to get the formula for the 
coupling $g_S$, 
\begin{equation} 
 g_S \simeq 
 \frac{30\, {\rm GeV}}{v_S} 
\times \left( \frac{m_{\Pi}}{750\,{\rm GeV}}  \right) 
\ll 1 
\,,  \label{gs-val}
\end{equation} 
which implies $v_S \gtrsim {\cal O}({\rm TeV})$.


\section{ The light pseudoscalar $s$ as a dark matter candidate} 
\label{s-DM}

As noted in the previous section, the present model predicts the light 
pseudoscalar $s$ as the direct consequence of the bosonic seesaw. 
In the present study we shall try to identify the $s$ as a dark matter candidate 
and this section devotes ourselves to discuss several cosmological and astrophysical limits on the $s$-dark matter.


\subsection{Lifetime}

We first evaluate the $s$ mass, decay property, and its lifetime.    
The $s$ mass is related to the HC pion masses through Eqs.(\ref{masses}) and (\ref{gs-val}) 
as 
\begin{equation} 
 m_s \simeq \sqrt{2} g_S \Lambda_{\rm HC} \simeq 42\, {\rm GeV} 
\times \left( \frac{m_{\Pi}}{750\,{\rm GeV}}  \right) 
\left( \frac{\Lambda_{\rm HC}}{1 {\rm TeV}} \right) 
\left( \frac{1 {\rm TeV}}{v_S} \right)
\,. \label{s-mass-val}
\end{equation} 
The $s$ couplings to the SM particles arise from mixing with the HC-eta prime coupled to 
the SM gauge bosons, $WW, ZZ, Z \gamma$ and $\gamma\gamma$, along with the tiny factor  
$g_S \ll 1$ (see Appendix~\ref{chiral-lag}). 
 Taking into account the size of the $s$ mass in Eq.(\ref{s-mass-val}), 
we thus find that the decay channel of the $s$ is only the diphoton mode through the vertex:
\begin{equation} 
 {\cal L}_{s \gamma\gamma} 
= - \frac{1}{4} g_{s\gamma\gamma} \, s \, F_{\mu\nu} \tilde{F}^{\mu\nu}
\,, \qquad 
\tilde{F}^{\mu\nu} \equiv \frac{1}{2}\epsilon^{\mu\nu\rho\sigma} F_{\rho\sigma}
\,, \label{int:s}
\end{equation}
with $F_{\mu\nu}=\partial_\mu A_\nu - \partial_\nu A_\mu$ and 
\begin{equation} 
 g_{s\gamma\gamma} 
= \frac{4 \sqrt{2}}{\pi} \sqrt{N_{\rm HC}} \frac{g_S \alpha_{\rm em}}{f} 
\simeq 
16 \sqrt{N_{\rm HC}} \alpha_{\rm em} \frac{m_s}{\Lambda_{\rm HC}^2} 
\,,  \label{gsgg}
\end{equation}
where use has been made of Eq.(\ref{s-mass-val}). 
The lifetime of $s$ is thus calculated to be 
\begin{eqnarray}
 \Gamma_s/N_{\rm HC} = \frac{g_{s\gamma\gamma}^2}{4096 \pi} m_s^3/N_{\rm HC} 
&\simeq& 
275 \,{\rm meV} 
\left( \frac{m_s}{42 \,{\rm GeV}} \right)^5
\left( \frac{1\,\rm TeV}{\Lambda_{\rm HC}} \right)^4 
\,. \label{lifetime-s} 
\end{eqnarray}

For the $s$ to be a dark matter, the lifetime has to be longer than the age of the universe 
at present time, which requires $\tau \gtrsim 10^{17}\, s$. 
From Eq.(\ref{lifetime-s}) the $s$ mass is thus constrained as 
\begin{equation} 
 m_s \lesssim 10 \,{\rm keV} \times 
 \left( \frac{\Lambda_{\rm HC}}{1 \, {\rm TeV}} \right)^{4/5}
\,. 
\end{equation}

\subsection{Astrophysical and cosmological limits}

\subsubsection{Line emission observations} 

\begin{figure}[t]
\begin{center}
   \includegraphics[width=9.0cm]{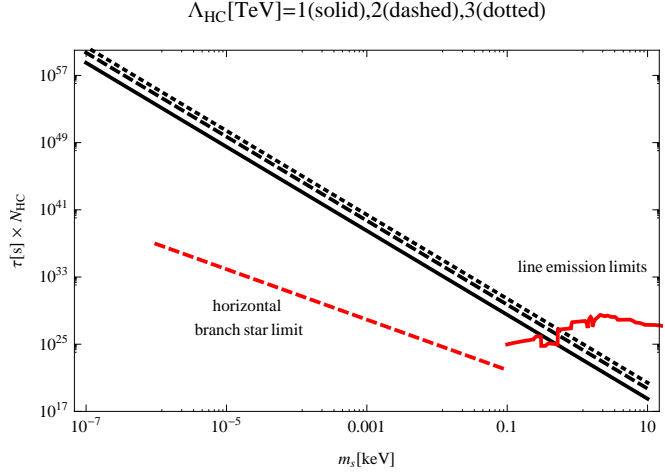} 
\caption{ 
The line emission and horizontal branch star observation limits on the $s$. 
The region below the red-solid and - dashed lines are excluded. 
The data have been quoted from Refs.~\cite{Baer:2014eja,Mambrini:2015sia,Raffelt:2006cw}.  
\label{limit-on-s-from-Xray}
}
\end{center} 
 \end{figure}

The $s$, dominantly decaying to photon,   
is expected to affect several line emission observations such as 
gamma-ray, X-ray, and cosmic ray,  
so the mass of $s$ can be severely constrained as 
in the case for other dark matter candidates~\cite{Mambrini:2015sia,Baer:2014eja}.  
In addition, 
the mass-independent limit 
on the coupling to the photon, $g_{s\gamma\gamma}$, 
can be placed by the observations of the horizontal branch stars 
for a lower mass range $m_s \lesssim 0.1$ keV~\cite{Raffelt:2006cw}.     
From Eq.(\ref{lifetime-s}) in Fig.~\ref{limit-on-s-from-Xray} 
we make a plot of the lifetime of $s$ ($\tau$) as 
a function of the mass $m_s$ in comparison with the 
line shape and the horizontal branch star limits. 
 The figure implies the limits on the $s$ as 
\begin{equation} 
 m_s \lesssim 1 \,{\rm keV} 
\,, \qquad 
{\rm with} 
\qquad  
\tau \times N_{\rm HC} \simeq 1.6 \times 10^{28}\,[s] 
\left( \frac{0.1 \,{\rm keV}}{m_s} \right)^5 
\left( \frac{\Lambda_{\rm HC}}{1\,\rm TeV} \right)^4 
\,.  
\label{X-ray-ms}
\end{equation}

\subsubsection{Constraints on the thermal $s$}

The $s$-dark matter can be thermally produced by the scattering with the photon, 
$s + \gamma \leftrightarrow s + \gamma$,  
through the interaction in Eq.(\ref{int:s}) with the coupling Eq.(\ref{gsgg})
in the early universe. The reaction rate $R(T)$ can roughly be estimated as 
\begin{equation} 
 R(T) = n(T) \langle \sigma v \rangle \approx 
g_{s\gamma\gamma}^4 T^5
\,.  
\end{equation}
The decoupling temperature of the $s$, $T_D$, 
can be evaluated, by equating this $R(T)$ with the Hubble rate $H(T) \sim\sqrt{ g_*(T)} T^2/M_P$ 
with the reduced Planck mass scale $M_P = 2.44 \times 10^{18}$ GeV and 
$g_*(T_D)$ being the effective degrees of freedom 
for relativistic particles. 
The $g_*(T_D)$ is estimated by combining the SM, the HC sector and the pseudoscalar 
$s$ as $g_*(T_D) = g_*^{\rm SM} + g_*^s + g_*^{\rm HC}$, 
where $g_*^{\rm SM}=106.75$~\cite{Agashe:2014kda} and $g_*^s =1$. 
The $g_*^{\rm HC}$ is calculated as 
\begin{eqnarray} 
g_*^{\rm HC} 
&=&
\left[2\times (N_{\rm HC}^2 - 1) \right]_{{\cal G}_{\rm HC}} 
+ \frac{7}{8} N_{\rm HC} \left[(2 \times 2 \times 2)_\chi + (2 \times 2)_\psi \right] 
\nonumber\\ 
&=& 
2 (N_{\rm HC}^2 -1 ) + \frac{21}{2} N_{\rm HC}
\, . 
\end{eqnarray}
For $N_{\rm HC}=(3,4,5)$, 
we have 
\begin{equation} 
g_*(T_D) = (155.25, 179.75, 208.25)
\,.
\end{equation}    
Thus we find 
\begin{equation}
 T_D \approx 10^{12}\,{\rm GeV} \times 
\left( \frac{3}{N_{\rm HC}} \right)^{2/3} 
\left( \frac{10^{-1}\,{\rm keV}}{m_s} \right)^{4/3} 
\left( \frac{\Lambda_{\rm HC}}{1 \,{\rm TeV}} \right)^{8/3} 
\left( \frac{g_*(T_D)}{200} \right)^{1/6} 
\,,\label{td}
\end{equation} 
where Eq.(\ref{gsgg}) have been used.

Even after decoupling from the thermal equilibrium, 
the $s$ (with mass $\lesssim 1$ keV as in Eq.(\ref{X-ray-ms})) can be    
still relativistic at present, 
which is constrained by the null observation of 
dark radiations~\cite{DiValentino:2016ikp}. 
Since the $s$ goes cool down just like radiations 
due to the Hubble expansion after the decoupling,  
the present temperature of the $s$ is estimated 
as 
\begin{equation} 
T_0(s) = (t_D/t_0)^{1/2} T_D = (g_*(T_0)/g_*(T_D))^{1/4} T_0
\simeq 10^{-4} {\rm eV} 
\,. 
\end{equation}  
with $g_*(T_0)= (2)_\gamma + (21/4(4/11)^{4/3})_{\nu} + (1)_s \simeq 4.36$.  
The current dark radiation constraint reads~\cite{DiValentino:2016ikp} 
$\Delta N_{\rm eff} = (T_0(s)/T_0(\nu))^3 < 0.1$ with $T_0(\nu)=(4/11)^{1/3}T_0$. 
The $s$ mass may thus  be required to be 
\begin{equation} 
 m_s \gtrsim 10^{-4}\,{\rm eV}
\,. \label{dr-ms} 
\end{equation}

If the $s$ decouples from the photon 
after the inflation and reheating temperature $T_R$, 
the temperature of the $s$ is heated back up to reach the same 
as the photon temperature, so that the $s$ would be a warm or hot dark matter-like particle. 
Currently such a light warm matter has been severely constrained by 
the cosmic microwave background spectrum. Hence we may escape 
from the case, by imposing $T_D > T_R$. 
The present model may follow a typical Higgs inflation scenario, as discussed 
in Ref.~\cite{Reheating}, in which $T_R \simeq 10^{14}$ GeV. 
 Taking this value as a reference and using Eq.(\ref{td}), we thus find 
\begin{equation} 
 m_s \lesssim 1\, {\rm eV} \times 
\left( \frac{3}{N_{\rm HC}} \right)^{1/2} 
\left( \frac{\Lambda_{\rm HC}}{1 \, {\rm TeV}} \right)^{2}
\left( \frac{g_*(T_D)}{200} \right)^{1/8}  
\,. \label{tr} 
\end{equation}

 From Eqs.(\ref{X-ray-ms}), (\ref{dr-ms}) and (\ref{tr}), 
we thus see the $s$ mass constrained to be 
\begin{equation} 
 10^{-4}\,{\rm eV} \lesssim m_s \lesssim 1\,{\rm eV} 
\,. \label{mass-range}
\end{equation}

\section{Cosmological productions and Detection of the $s$-dark matter} 
\label{prod-s}

In this section, we closely explore the possibility for the $s$ as a dark matter 
to account for the relic abundance at the present time.

\subsection{Thermal production} 

Though the $s$-dark matter decouples from the thermal equilibrium in the early universe 
at $T_D \approx 10^{14}$ GeV $\times (1\,{\rm eV}/m_s)^{4/3} (\gtrsim T_R)$, 
there might exist the chance to thermally accumulate the number density by production cross sections 
interacting with the HC sector until the HC sector decouples from the thermal equilibrium 
at around $T=\Lambda_{\rm HC}={\cal O}({\rm TeV})$.   
The relevant production processes involve only a single $s$ in the final state through 
the $s-\gamma-\gamma$ vertex in Eq.(\ref{int:s})\footnote{When the temperature is significantly higher than $\Lambda_{\rm HC}$, the $s$-coupling to diphoton may arise from the HC fermion loops.  Even if the universe is in such a symmetric phase by taking into account the thermal effect, the vertex is anyhow generated with the magnitude of the order of $g_{s \gamma \gamma}$ which is given by Eq.~(\ref{gsgg}).} and the $s-Z-\gamma$, $s-Z-Z$ vertices 
listed in Appendix~\ref{chiral-lag}, scattered off from the HC sector-fermion $F=(\chi, \psi)$ 
such as  
$ F + \bar{F} \to \gamma/Z + s $.  
 The production cross section roughly goes like 
\begin{equation}  
\sigma(F +\bar{F} \to \gamma/Z + s) 
\sim \alpha_{\rm em} N_{\rm HC} 
\left( \frac{\sqrt{N_{\rm HC}} g_S \alpha_{\rm em}}{ \Lambda_{\rm HC}} \right)^2 
\simeq 
10^{-31} \times \frac{N_{\rm HC}^2}{\Lambda_{\rm HC}^2} \left( \frac{m_s}{1\,{\rm eV}} \right)^2 \left( \frac{1\,{\rm TeV}}{\Lambda_{\rm HC}} \right)^2
\,, \label{cross-s-HC}
\end{equation}  
where in the second equality we have used the first relationship in Eq.(\ref{s-mass-val}). 
The corresponding number density per entropy density at present time ($Y_s(T_0)=n_s(T_0)/s(T_0)$)
can be estimated by integrating the Boltzmann equation with the above production cross section  
over the temperature from the reheating temperature $T_R\approx 10^{14}$ GeV 
down to the freeze-out temperature $T_F=\Lambda_{\rm HC}$. 
Following the formula given in Ref.~\cite{Choi:1999xm} 
we thus evaluate the $Y_s(T_0)$ as 
\begin{eqnarray} 
  Y_s(T_0) 
  &=& \int_{\Lambda_{\rm HC}}^{T_R} dT \frac{\langle \sigma (F + \bar{F} \to \gamma/Z + s) v \rangle n_F n_{\bar{F}} }{s(T) H(T) T}
\, \nonumber \\ 
 &=& 
 \frac{135\sqrt{10} M_P}{2 \pi^{3}} \int_{\Lambda_{\rm HC}}^{T_R} dT \frac{\langle \sigma  (F + \bar{F} \to \gamma/Z + s) v  \rangle n_F n_{\bar{F}}}
{g_{*}^{3/2} (T) T^6}
 \,, 
\end{eqnarray}
where in reaching the last line we used 
$H^2(T)=\frac{\pi^2}{30} g_*(T) T^4/(3M_P^2)$, 
$s(T)=g_{s*}(T) \frac{2 \pi^2}{45} T^3$, with $g_*(T) = g_{s*}(T)$ is assumed 
and the thermal average is expressed to be 
\begin{equation} 
  \langle \sigma_n (F + \bar{F}\to s + \gamma/Z) v \rangle n_F n_{\bar{F}} 
  =  \zeta^2(3) \cdot \eta_F \eta_{\bar{F}}  \cdot \frac{g_Fg_{\bar{F}}}{16\pi^4} T^6 \int_0^\infty dx x^4 K_1(x) \sigma(x^2) 
  \,, 
\end{equation} 
where $\zeta(3)= 1.202...$ 
and $K_1(x)$ stands for the modified Bessel function of the first kind, 
$\sigma(x^2) = \sigma(s/T^2)$ and $g_{F(\bar{F})}$ is the internal (spin) degree of freedom 
for the HC fermion (anti-fermion) $F$; $\eta_{F(\bar{F})}$ is a number density factor associated with the initial state particle   
assigned as $\eta_F=3/4$ for fermions (anti-fermion). 
Using these we thus calculate the $Y_s(T_0)$ to get 
\begin{eqnarray} 
 Y_s(T_0) 
&\approx &
 \frac{135 \sqrt{10}}{32 \pi^6} \frac{M_P T_R}{g_*^{3/2}(T_R) \Lambda_{\rm HC}^2} \times 10^{-31} \times N_{\rm HC}^2 
 \left( \frac{m_s}{1\,{\rm eV}} \right)^2 \left( \frac{1\,{\rm TeV}}{\Lambda_{\rm HC}} \right)^2 
 \nonumber \\ 
 &\approx& 
 10^{-10} \times N_{\rm HC}^2  \left( \frac{m_s}{1\,{\rm eV}} \right)^2 \left( \frac{1\,{\rm TeV}}{\Lambda_{\rm HC}} \right)^4 
 \left(\frac{200}{g_*(T_R)}\right)^{3/2} 
 \,, \label{Eq:Y_th}
\end{eqnarray} 
where use has been made of $g_*(T_R)=g_*(\Lambda_{\rm HC})$. 
Thus, it turns out that the thermal relic is too small to explain the present dark matter abundance.  
This result is essentially tiled with the tiny coupling $g_S$ which leads to the extremely 
small cross section with the HC sector in Eq.(\ref{cross-s-HC}).

\subsection{Non-thermal production} 

Analogously to the case of axion dark matter~\cite{Baer:2014eja}, 
the $s$-dark matter population can be accumulated by ``misalignment" of
the classical $s$ field and the coherent oscillation. 
Assuming the initial position at which the oscillation starts to be 
the vicinity of the vacuum $s=0$ with the vacuum expectation value $v_S$, 
we write 
the equation of motion for the $s$ under the Friedmann-Robertson-Walker metric 
to be 
\begin{equation} 
 \frac{d^2 s}{dt^2} 
 + 3 H(T) \frac{d s}{dt} + m_s^2 \, s \approx 0 
 \,. 
\end{equation}
This describes the damping harmonic oscillation in which the oscillation 
takes place when $T=T_{\rm osc}$ where $3 H(T) \approx m_s $, i.e., 
\begin{equation}
 T_{\rm osc} 
 \simeq 
 13 \,{\rm TeV} \times \left( \frac{m_s}{1 \, \rm eV} \right)^{1/2} \left(\frac{ 200}{g_*(T_{\rm osc})} \right)^{1/4} 
 \,. 
\end{equation}
This implies that $130\,{\rm GeV} \lesssim T_{\rm osc} \lesssim 13$ TeV for $10^{-4}\,{\rm eV} \lesssim m_s \lesssim 1$ eV. 
Since the $s$ mass is generated through the bosonic seesaw at $T\simeq \Lambda_{\rm HC}={\cal O}(1)$ TeV, 
we find that the temperature at which the coherent oscillation starts, what we call $T_S$, depend on the $m_s$ as 
\begin{eqnarray} 
 T_S &\simeq& \Lambda_{\rm HC} \,\qquad {\rm for} \qquad  6 \times10^{-3} \,{\rm eV} \left( \frac{\Lambda_{\rm HC}}{1\,{\rm TeV}} \right)^2 \lesssim m_s < 1\,{\rm eV}
 \,, \nonumber \\ 
 T_S &\simeq& T_{\rm osc} \, \qquad {\rm for} \qquad 10^{-4}\,{\rm eV} \lesssim m_s \lesssim 6 \times10^{-3} \,{\rm eV} \left( \frac{\Lambda_{\rm HC}}{1\,{\rm TeV}} \right)^2
 \,. \label{TS}
\end{eqnarray}
The energy density of the classical $s$ field is thus accumulated by the coherent oscillation 
starting from the temperature $T_S$ in Eq.(\ref{TS}), cooling down to the present temperature $T_0$.

 At the $T=T_S$ the energy density of the $s$ corresponds to the vacuum energy defined as 
 \begin{equation} 
  \rho_s(T_S) = V(\theta) - V(\theta=0)
  \,, 
 \end{equation}
where the $\theta$ is defined as the amount of the shift from the original $S$ field at the vacuum expectation value $v_S$ 
to be $S=v_S(1 + \theta)$ with $\theta \ll 1$, and the potential $V(\theta)$ is read off as 
\begin{equation} 
 V(\theta) = V(\theta=0) + \frac{1}{2} m_s^2 v_s^2 \theta^2 + {\cal O} (\theta^3) 
\,.  \label{V-theta}
\end{equation}
One can easily see that during the coherent oscillation, the number density per comoving volume 
is conserved and the $s$ behaves just like a non-relativistic particle satisfying $\rho_s \propto R^{-3}$ 
with the expansion rate $R$. Hence we write  
\begin{equation} 
 \frac{\rho_{s}(T_S)}{\rho_s(T_0)} 
 = \frac{m_s n_s(T_S)}{m_s n_s(T_0)} 
 = \frac{s(T_S)}{s(T_0)} 
 \,, 
\qquad {\rm i.e.,} \qquad  
\rho_s(T_0) = \frac{s(T_0)}{s(T_S)} \rho_s(T_S) 
\,.  \label{rhoS}
\end{equation}
Thus, we get the present abundance of DM as 
\begin{eqnarray}
\rho_s (T_0) 
\simeq 
(4200~{\rm GeV})^4
\left( 
\frac{T_0}{T_S} 
\right)^3 
\frac{g_{\ast S}(T_0)}{g_{\ast S}(T_S)} \theta^2 
\left( 
\frac{m_\Pi}{750~{\rm GeV}}
\right)^2 
\left(
\frac{\Lambda_{\rm HC}}{1~{\rm TeV}}
\right), \label{srelic}
\end{eqnarray}
by using Eqs.~(\ref{s-mass-val}).
This relation shows that we can explain the correct abundance of $s$ with an appropriate value of $\theta$ 
even when the HC pion mass is heavier/lighter than $750$ GeV.

\begin{figure}[t]
\begin{center}
   \includegraphics[width=7.0cm]{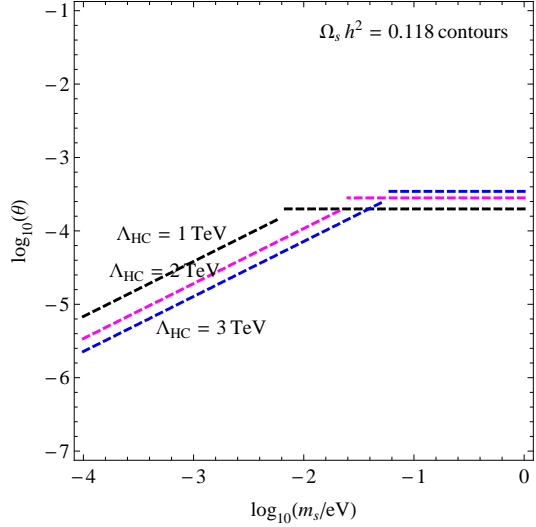} 
\caption{ 
The contour plots on the $(m_s, \theta)$ plane realizing the 
observed present dark matter density $\Omega_{\rm DM}h^2 =0.118$~\cite{Agashe:2014kda}. 
The bumps, which show up when $T_S$ gets lower than $\Lambda_{\rm HC}$, are due to 
the discontinuity of the effective degrees of freedom $g_{s*}$ around the $T_S = \Lambda_{\rm HC}$ 
as described in the text.  
\label{Coherent-Osc-s}
}
\end{center} 
 \end{figure}

From Eqs.(\ref{TS}), (\ref{V-theta}) and (\ref{rhoS}), and using the second equality in Eq.(\ref{s-mass-val}), 
we thus estimate the $s$-dark mater relic density, $\Omega_s h^2 = \rho_s(T_0)/(\rho_{\rm cr}/h^2)$ 
with $\rho_{\rm cr}/h^2 = 0.8 \times 10^{-46}\,{\rm GeV}^4$. 
The contour plot on the $(m_s, \theta)$ plane with the 
observed dark matter relic density $\Omega_{\rm DM} h^2 \simeq 0.118$~\cite{Agashe:2014kda} 
has been drawn in Fig.~\ref{Coherent-Osc-s}. 
Here use has been made of 
$s(T_0) = \frac{2 \pi^2}{45} g_{s*}(T_0) T_0^3$ with $g_{*s}(T_0)=43/11$ and $T_0\simeq 2.4 \times 10^{-4}$ eV,  
$g_{s*}(\Lambda_{\rm HC})=200$ taken as a reference value, 
and we have assumed $g_{s*}(T_S < \Lambda_{\rm HC}) = g_*^{\rm SM}=106.75$. 
 From the figure, we find that 
the relic density of the $s$, with the mass in a range of $10^{-4} \, {\rm eV} \lesssim m_s 
\lesssim 1\,{\rm eV}$,  
can be accumulated enough to account 
for the present dark matter abundance.

\subsection{Detection possibility in experiments}

As has so far been seen in this section, 
the $s$-dark matter has the lifetime much longer than the age of the universe 
and has extremely tiny couplings to the SM particles,  
and hence the detection at collider experiments is unlikely to be possible.

As in the case of invisible axion-like dark matter detection~\cite{SikivieDetection},  
cosmic pseudoscalar $s$, left over from the big bang, 
may be detected by microwave cavity haloscopes.  
In that facility, a strong static magnetic field is provided 
to make the $s$ drift through the microwave cavity, resonantly converted to 
microwave photons according to the $s$-photon-photon interaction in Eq.(\ref{int:s}). 
The conversion power $P$ is given by~\cite{SikivieDetection}
\begin{equation} 
P = \frac{1}{8} g_{s\gamma\gamma}^2 \rho_s(T_0)B_0^2 L_x V 
\,, 
\end{equation}
where $\rho_s(T^0)$ is the local $s$ energy density, $B_0$ the magnetic strength scale, 
$V$ the volume of the cavity and $L_x$ the size of the $x$ direction. 
Taking a typical experimental setup currently employed~\cite{Bradley:2003kg}, 
$B_0=10\, {\rm Tesla}, $ $L_x=1\, {\rm m}$, $V=1\,{\rm m}^3$ 
and the local halo density 
$\rho_{\rm halo}\simeq0.3\, {\rm GeV}/{\rm cm}^3$,  
  we estimate the detection power 
 \begin{equation} 
 P/N_{\rm HC} \simeq 10^{-34} \,{\rm Watt} \times 
\left( \frac{1 \,{\rm TeV}}{\Lambda_{\rm HC}} \right)^2 
\left( \frac{m_s}{10^{-4} \,{\rm eV}}  \right)^2 
\left( \frac{\rho_s(T_0)}{\rho_{\rm halo}} \right)
\,, \label{power:s}
\end{equation} 
where we have used Eq.(\ref{gsgg}). 
The power for $m_s \sim 1$ eV  
 is comparable with the axion detection potential~\cite{Bradley:2003kg}, so 
the $s$ can be hunted at the same level of the sensitivity 
as the axion by the microwave cavity experiments.

\section{Summary and Discussion} 
\label{summary}

In this paper, 
we have employed a model based on
the classically-scale invariant standard model extended by adding 
a strongly coupled hypercolor dynamics.
The dynamical breaking of the scale symmetry is triggered 
by the vector-like condensation at the TeV scale, 
so that the standard model Higgs acquires the negative mass-squared by the bosonic seesaw mechanism
to realize the electroweak symmetry breaking.

What is significant to control this model is to 
include an elementary pseudoscalar $S$, which plays the crucial role 
to realize the electroweak bosonic seesaw, as well as  
to give masses for the composite Nambu-Goldstone bosons (hypercolor pions):  
in this sense, the $S$ acts like another ``Higgs'' in the theory. 
Thus, discovering the fluctuating mode of $S$, called $s$, is the smoking-gun of the present 
model.

Because of the classical-scale invariance, the pseudoscalar $S$ originally 
couples only to the standard model Higgs and hypercolor fermions. 
After the dynamical-scale breaking and triggering the electroweak bosonic seesaw,  
the $s$ thus develops vanishingly small couplings to the standard model particles, which arise only 
through the tiny mixing with the hypercolor eta-prime. 
In addition, it turned out that in relation to the hypercolor pion masses 
the $s$ mass is predicted to be very light and to predominantly decay to diphoton, 
so we have identified the $s$ as a dark matter candidate.   
The $s$ mass was then severely 
constrained by several cosmological observations, such as line emissions 
of X-ray, gamma-ray and 
cosmic-ray, no evidence for dark radiations, and a typical Higgs inflation scenario. 
The $s$ mass was thus bounded to be $10^{-4} {\rm eV} \lesssim m_s \lesssim 1 {\rm eV}$.

We examined the possibility of the cosmological productions of the $s$. 
It was shown that the $s$ is unlikely to be thermally produced essentially 
due to its tiny couplings to the hypercolor sector in the thermal equilibrium.  
We then found that the sufficient amount of relic abundance of $s$ as the cold dark matter 
can be accumulated via the coherent oscillation.

The detection potential in microwave cavity experiments was also addressed 
so that the $s$ with mass around 1 eV can have the same level of the 
detection sensibility as that of the axion in the currently equipped experimental 
setup, so the $s$ can be hunted by the microwave cavity experiments.  
\\

Several comments are in order: 
\\

The crucial deference between the $s$-dark matter and the axion-like dark matter 
can be seen by no evidence for observations probing couplings to 
matter, such as the test of gravitational inverse-square law and energy loss in stars 
like neutron star cooling. 
The $s$ coupling to matters can be generated at loop levels 
by the $g_S$ and $\kappa_H$ couplings. 
As seen from Eqs.(\ref{stationary:condi}) and (\ref{masses}), however, 
those couplings are extremely small, suppressed by $(m_s/\Lambda) \ll 1$ or 
$(v_1/v_s) \ll 1$ (See also Eq.(\ref{gs-val})). 
Hence one can conclude that there is no chance to detect the $s$-dark matter 
through the couplings to matters, in contrast to the axion case. 
Thus, no evidence for observations with the matter-portal, but 
some signals identical among the $s$ and the axion 
in the line shapes and microwave cavity experiments 
would be a clear hint to distinguish them. 
(Note that a dilaton-like dark matter signal in the microwave cavity 
is clearly different from that of the $s$ and the axion, due to 
the different type of the coupling to photons: $E\cdot B$ for pseudoscalars, while 
$E\cdot E$ or $B\cdot B$ for scalars.)

As discussed in Secs.~\ref{s-DM} and \ref{prod-s},  
we have assumed that 
the reheating epoch is associated with the Higgs inflation scenario. 
It might be the case, however, that 
one needs somewhat large  
non-minimal couplings  
between the SM Higgs and the scalar curvature for 
the reheating temperature in the Higgs inflation scenario. 
In that case, the reheating epoch would be shifted, 
so the upper bound on the mass of $s$,  as estimated in Eq.~(\ref{tr}), 
could be affected. 
Detailed study closely connected with inflation scenarios  
is to be performed in the future literature.

The predicted number in Eq.(\ref{power:s}) depends on 
the $s$ mass, so it does also on the hypercolor pion mass through Eq.(\ref{s-mass-val}). 
It should be noted, however, that the light pseudoscalar $s$ as a candidate of 
the dark matter is intact even if the hypercolor pion mass is not set to the present reference value, 
since it is solely tied with realization of the electroweak breaking via the 
bosonic seesaw: the mass has to be much smaller than $\Lambda_{\rm HC}$, 
which is controlled by the small coupling $g_S(\ll 1)$ in Eq.(\ref{masses}); 
the $s$ couplings to the standard model particles, photons, necessarily becomes tiny by the same 
$g_S$ coupling strength as the consequence of the bosonic seesaw, 
which would suggest to regard the $s$ as a dark matter candidate; 
the $s$ mass is then 
inevitably constrained by cosmological bounds, to be order of eV, as was discussed 
in the text. 

Actually, the couplings of the s-dark matter are required to be extremely small:
the coupling to hyperfermion, $g_S \sim 10^{-12}$, from Eq.~(\ref{s-mass-val}) for $m_s \sim 1 {\rm eV}$ 
(which is coincidentally as small as the Yukawa coupling for neutrino in Dirac neutrino models);
the quartic coupling $\lambda_S \sim 10^{-44}$ from Eq.~(\ref{masses}) with $v_S \sim 10^{13}$ GeV
estimated from Eq.~(\ref{gs-val})  with $m_s \sim 1 {\rm eV}$ and $g_S \sim 10^{-12}$;
the coupling to the 125 GeV Higgs, $\kappa_H \sim 10^{-22}$, estimated from Eq.~(\ref{masses}) with
$v_S \sim 10^{13}$ GeV. The origin of these extremely small couplings could be explained by
the underlying Planck scale physics, which is, however, beyond the scope of the present study,
to be pursued elsewhere. Note that the realization of the electroweak symmetry breaking has nothing
theoretically to do with the smallness of those coupling parameters, which are only related to
the physics of the light $s$ including the mass generation of hypercolor pions and the property as
the invisible dark matter.

Other signals characteristic to the present model involve 
not only hypercolor pions, but also the hypercolor eta-prime and 
hypercolor composite scalar states, 
both of which are expected to have the mass on the order of $\Lambda_{\rm HC}$.    
As briefly studied in Appendix~\ref{750}, the hypercolor eta-prime can be produced 
at the LHC, via the photon - photon fusion process as well as the hypercolor pions. 
The discovery channels will be similar to the hypercolor pions: $WW, ZZ, Z\gamma$ and 
$\gamma\gamma$ modes. Since the production cross section decreases as the 
resonance mass grows, the photon - photon fusion cross section for the 
hypercolor eta-prime significantly gets smaller than that of the hypercolor pions, 
so it may be challenging to search at the LHC (For explicit estimates for 
the signal strengths, see Appendix~\ref{750}).    

As to the hypercolor composite scalars, the couplings to the standard model particles 
are controlled by the tiny Yukawa coupling $y(\ll 1)$ through 
the mixing with the standard model Higgs. It would be worth investigating how 
much large the $y$ coupling is allowed to be consistent with the 
currently reported heavy Higgs search data, and to discuss the 
LHC discovery potential. Such those topics are deserved to the future 
study.

In closing, in the present work we have so far focused on the possibility 
for the predicted light pseudoscalar $s$ to be a dark matter candidate. 
Actually, another scenario can be made: with the $s$ mass around GeV scale 
the $s$ could be just a long-lived particle having the lifetime much 
shorter than the age of the present universe. 
That sort of a light long-lived particle could be 
accessible at the LHC. This interesting another 
possibility will be pursued in another publication.


\acknowledgments

This work was supported in part by 
the JSPS Grant-in-Aid for Young Scientists (B) \#15K17645 (S.M.) 
and Research Fellowships of the Japan Society for the Promotion of Science for Young Scientists
 \#26$\cdot$2428 (Y.Y.).

\appendix 

\section{Computation of HC pion masses}
\label{Pi-mass}

In this Appendix we shall calculate the HC pion masses 
arising from the $g_S$ and $y$ terms in Eqs.(\ref{gS-term}) and (\ref{y}).

\subsection{Masses from the $g_S$ term}

First of all, one should note that the nonzero vacuum expectation value of $S$, $v_S$, 
is required in the present model, which  
provides 
masses for the 8 HC pions $(\Pi^a)$ via the $g_S$ term in Eq.(\ref{gS-term}). 
The HC pion masses can be evaluated 
according to the standard current algebra, 
which turn out to show up at the second order of perturbation in $g_S$: 
\begin{equation} 
(m_\Pi^2)^{ab} \Bigg|_{g_S}
= - \frac{i}{2} g_S^2 v_S^2 \int d^4x \langle \Pi^a | T(J_P(x)J_P(0)) |\Pi^b \rangle 
\,, \label{DF}
\end{equation} 
where $J_P(x) = i \bar{F}(x) \gamma_5 F(x)$ and the symbol ``$T$'' stands for the time-ordered product.  
We use the partially-conserved axialvector current (PCAC) 
relations and the current algebra,  
\begin{eqnarray} 
&& \partial^\mu J_{\mu 5}^a(x) = - f_\Pi m_\Pi^2 \, \Pi^a(x) 
\,, \nonumber \\ 
&& [i Q_5^a, {\cal O}(x)] = \delta^a_5 {\cal O}(x)
\,, \qquad 
Q_5^a = \int d^3 x J_5^0(x)\,, 
\label{PCAC}
\end{eqnarray} 
where the current $J_{\mu 5}^a$ is defined as 
$J_{\mu 5}^a = \bar{F} \gamma_\mu \gamma_5 (\lambda^a/2) F$ 
with the generator $(\lambda^a/2)$ with the Gell-Mann matrix $\lambda^a$ ($a=1,\cdots 8$); 
$f_\Pi$ is the $\Pi$-decay constant, defined as 
$\langle 0| J_\mu^a(0) | \Pi^b(p) \rangle = - i p_\mu f_\Pi \delta^{ab}$; 
the ${\cal O}$ denotes an arbitrary Heisenberg operator, and 
$\delta^a_5$ denotes the infinitesimal-``chiral'' transformation, 
which acts on the $F$-fermion as 
$\delta^a_5 F = - i \gamma_5 (\lambda^a/2) F$. 
Using these together with the reduction formula, 
one thus evaluates Eq.(\ref{DF}) to arrive at  
\begin{eqnarray}
 (m_\Pi^2)^{ab} \Bigg|_{g_S}
&=& m_\Pi^2 \, 
=  - 4 i \frac{g_S^2 v_S^2}{f_\Pi^2} \, \delta^{ab} \,  
\int d^4 x 
\left( 
 \langle 0|T (J_S^a(x) J_S^b(0)) |0 \rangle 
-  
 \langle 0|T (J_{\eta'}(x) J_{\eta'}(0)) |0 \rangle \delta^{ab} 
\right)  
\nonumber \\ 
&=& 
  4  \frac{g_S^2 v_S^2}{f_\Pi^2} \, \delta^{ab} \,  
\left[ 
 \Pi_{S}(0) -  \Pi_{\eta'}(0) 
\right]
\,, \label{mass:formula} 
\end{eqnarray}
where $J_S^a(x) = \bar{F}(x) (\lambda^a/2) F(x), J_{\eta'}(x)= \frac{1}{\sqrt{6}} \bar{F}(x) i \gamma_5 F(x)
$ and we have defined the current correlators 
$\Pi_{S,\eta'}$ as 
\begin{eqnarray} 
\,   \int d^4 x e^{i px} 
 \langle 0|T (J_{S}^a(x) J_{S}^b(0)) |0 \rangle 
&\equiv& 
i \Pi_{S}(p^2) \delta^{ab} 
\, \nonumber \\ 
\,   \int d^4 x e^{i px} 
 \langle 0|T (J_{\eta'}(x) J_{\eta'}(0)) |0 \rangle 
&\equiv& 
i \Pi_{\eta'}(p^2) 
\,.     
\end{eqnarray}
We may expand the correlators by assuming the resonances pole saturation, 
\begin{eqnarray} 
\Pi_{S}(p^2) 
&=& \sum_{n=1}^\infty 
 \frac{F_{S_n}^2 m_{S_n}^2}{ m_{S_n}^2 - p ^2} 
\,, \nonumber \\ 
\Pi_{\eta'}(p^2) 
&=& \sum_{n=1}^\infty 
 \frac{F_{\eta_n'}^2 m_{\eta_n'}^2}{ m_{\eta'_n}^2 - p ^2} 
\,,  
\end{eqnarray}
with the masses $(m_{S_n}, m_{\eta'_n})$ and the decay constants $(F_{S_n}, F_{\eta'_n})$.  
Then the HC pion mass formula in Eq.(\ref{mass:formula}) is rewritten as a sum rule 
to be 
\begin{eqnarray}  
m_\Pi^2\Bigg|_{g_S} = 4  \frac{g_S^2 v_S^2}{f_\Pi^2} 
\sum_n 
\left[ 
F_{S_n}^2 - F_{\eta_n'}^2  
\right]
\,. 
\end{eqnarray}
Analogously to  the QCD case,  
 the $\eta'\equiv \eta'_{(n=1)}$ decay constant $F_{\eta'_1}$ is expected to be of order of 
the pion decay constant~\cite{Gasser:1984gg}, $f_\Pi \sim {\cal O }( \frac{\Lambda_{\rm HC}}{4\pi})$,  
and the higher resonance contributions could numerically be cancelled each other in the sum 
between the scalar and pseudoscalar sector, namely, by $F_{S_n} \simeq F_{\eta'_n} \simeq {\cal O}(\Lambda_{\rm HC})$ for $n \ge 2$.  
Thus we may evaluate the sum rule just by keeping the lowest resonance contribution: 
\begin{eqnarray}  
m_\Pi^2 \Bigg|_{g_S}
&\simeq& 4  \frac{g_S^2 v_S^2}{f_\Pi^2} 
\left( 
F_{S_1}^2 - F_{\eta'_1}^2  
\right) 
\nonumber \\ 
&\simeq& 
 4  \frac{g_S^2 v_S^2}{f_\Pi^2/(N_{\rm HC}/3)} 
\Lambda_{\rm HC}^2  
\,, \label{approx:pNGmass}
\end{eqnarray} 
where in the last line we have clarified that the mass is independent of 
the number of HC, $N_{\rm HC}$. 

Thus the derivation of Eq.(\ref{HC-pion-mass}) has been compensated.

\subsection{Masses from the $y$-term}

Similarly to the $g_S$ term, the $y$-Yukawa term (${\cal L}_y$) in Eq.(\ref{y}) 
gives masses to the  HC pions via the $H$-Higgs vacuum expectation value $v_1 \simeq 246$ GeV. 
Again, the estimate of the mass can be done by using the current algebra technique: 
\begin{eqnarray} 
 (m_\Pi^2)^{ab}\Bigg|_{y} 
&=& - \frac{1}{f_\Pi^2} \langle 0| [i Q_5^a, [i Q_5^b, {\cal L}_y]] |0 \rangle 
\nonumber \\ 
&=& 
- \frac{y v_1}{\sqrt{2} f_\Pi^2} \langle 0| [i Q_5^a, [i Q_5^b, \bar{\chi}_2 \psi + \bar{\psi} \chi_2]] |0 \rangle 
\,. 
\end{eqnarray}
The nonzero elements for the mass matrix are thus found be 
\begin{eqnarray} 
  (m_\Pi^2)^{14}\Bigg|_{y} &=& 
(m_\Pi^2)^{14}\Bigg|_{y} = 
- \frac{y v_1}{f_\Pi^2} 
\langle \bar{F}F \rangle \,, \qquad 
  (m_\Pi^2)^{25}\Bigg|_{y} = 
(m_\Pi^2)^{52}\Bigg|_{y} = 
- \frac{y v_1}{f_\Pi^2} 
\langle \bar{F}F \rangle \,, \nonumber \\  
  (m_\Pi^2)^{36}\Bigg|_{y} &=& 
(m_\Pi^2)^{63}\Bigg|_{y} = 
 \frac{y v_1}{f_\Pi^2} 
\langle \bar{F}F \rangle \,, \qquad 
  (m_\Pi^2)^{68}\Bigg|_{y} = 
(m_\Pi^2)^{86}\Bigg|_{y} = 
\sqrt{3} \frac{y v_1}{f_\Pi^2} 
\langle \bar{F}F \rangle \,,   
\label{y-mass} 
\end{eqnarray}
where $\langle \bar{F}F \rangle$ denotes the ``chiral'' condensate per flavors, i.e.,  
$\langle \bar{F}F \rangle = \langle \bar{\chi}_1 \chi_1  \rangle
= \langle \bar{\chi}_2 \chi_2  \rangle = \langle \bar{\psi} \psi  \rangle$.

\subsection{Diagonalization of the HC pion sector}

Combining Eqs.(\ref{approx:pNGmass}) with Eq.(\ref{y-mass}).
 one finds the HC pion mass matrix acting on the current-eigenstate 
vector $(\Pi^1, \cdots , \Pi^8)^T$: 
\begin{eqnarray} 
\left( 
\begin{array}{c|c|c|c|c|c|c|c}
m^2_{g_S} & 0 & 0& (m^2_{y})^{14} & 0 & 0 & 0 & 0 \\ 
\hline 
0 & m^2_{g_S} & 0 & 0 & (m^2_{y})^{25} & 0 & 0 & 0 \\ 
\hline  
0 & 0 & m^2_{g_S} & 0 & 0 & (m^2_y)^{36} & 0 & 0 \\ 
\hline 
(m^2_{y})^{41} & 0  & 0 & m^2_{g_S} & 0 & 0 & 0 & 0 \\ 
\hline 
 0 & (m^2_{y})^{52} & 0 & 0 & m^2_{g_S} & 0 & 0 & 0 \\ 
\hline 
0 & 0 & (m^2_{y})^{63} & 0 & 0 & m^2_{g_S} & 0 & (m^2_{y})^{68} \\ 
\hline 
0 & 0 & 0 & 0 & 0 & 0 & m^2_{g_S} & 0 \\ 
\hline 
0 & 0 & 0 & 0 & 0& (m^2_{y})^{86} & 0 & m^2_{g_S} 
\\ 
\end{array}
\right)
\,,  
\end{eqnarray}
where $m^2_{g_S}$ and $(m_y^2)^{ab}$ respectively stand for 
the masses in Eqs.(\ref{approx:pNGmass}) and (\ref{y-mass}). 
The mass matrix can easily be diagonalized by an orthogonal rotation, 
which relates the current eigenstates $\{ \Pi \}$ with the mass eigenstates $\{ \tilde{\Pi} \}$ 
as
\begin{eqnarray} 
\left( 
\begin{array}{cc} 
\tilde{\Pi}^1 \\ 
\tilde{\Pi}^4
\end{array}
\right)
&=& 
\left( 
\begin{array}{cc} 
- \frac{1}{\sqrt{2}} & \frac{1}{\sqrt{2}} \\ 
\frac{1}{\sqrt{2}} & \frac{1}{\sqrt{2}} 
\end{array}
\right)
\left( 
\begin{array}{cc} 
{\Pi}^1 \\ 
{\Pi}^4
\end{array}
\right)
\,, \nonumber \\ 
\left( 
\begin{array}{cc} 
\tilde{\Pi}^2 \\ 
\tilde{\Pi}^5
\end{array}
\right)
&=& 
\left( 
\begin{array}{cc} 
- \frac{1}{\sqrt{2}} & \frac{1}{\sqrt{2}} \\ 
\frac{1}{\sqrt{2}} & \frac{1}{\sqrt{2}} 
\end{array}
\right)
\left( 
\begin{array}{cc} 
{\Pi}^2 \\ 
{\Pi}^5
\end{array}
\right)
\,, \nonumber \\ 
\left( 
\begin{array}{cc} 
\tilde{\Pi}^3 \\ 
\tilde{\Pi}^6 \\ 
\tilde{\Pi}^8
\end{array}
\right)
&=& 
\left( 
\begin{array}{ccc} 
- \frac{\sqrt{3}}{\sqrt{2}} &0 & \frac{1}{2} \\ 
\frac{1}{2 \sqrt{2}}  & - \frac{1}{\sqrt{2}} & \frac{1}{2} \sqrt{\frac{3}{2}} \\
\frac{1}{2 \sqrt{2}}  &  \frac{1}{\sqrt{2}} & \frac{1}{2} \sqrt{\frac{3}{2}} 
\end{array}
\right)
\left( 
\begin{array}{cc} 
{\Pi}^3 \\ 
{\Pi}^6 \\ 
{\Pi}^8
\end{array}
\right)
\,, \nonumber \\ 
\tilde{\Pi}^7 &=& \Pi^7 
\,, \label{mass-eigenstates:pi}
\end{eqnarray}
with the mass eigenvalues, 
\begin{eqnarray} 
m_{\tilde{\Pi}^1}^2 &=& 
m_{\tilde{\Pi}^2}^2  
\simeq m_{g_S}^2 + \frac{y v_1 \langle \bar{F}F \rangle}{\sqrt{2} f_\Pi^2} 
\,, \nonumber \\ 
m_{\tilde{\Pi}^3}^2 &\simeq& m_{g_S}^2 
\,, \nonumber \\  
m_{\tilde{\Pi}^4}^2 &=& 
m_{\tilde{\Pi}^5}^2  
\simeq m_{g_S}^2 - \frac{y v_1 \langle \bar{F}F \rangle}{\sqrt{2} f_\Pi^2} 
\,, \nonumber \\ 
m_{\tilde{\Pi}^6}^2 &\simeq& 
m_{g_S}^2 - \frac{\sqrt{2}y v_1 \langle \bar{F}F \rangle}{f_\Pi^2} 
\,, \nonumber \\ 
m_{\tilde{\Pi}^7}^2 &\simeq& m_{g_S}^2 
\,, \nonumber \\ 
m_{\tilde{\Pi}^8}^2 &\simeq& 
m_{g_S}^2 + \frac{\sqrt{2}y v_1 \langle \bar{F}F \rangle}{f_\Pi^2} 
\,,  \label{pNGmasses}
\end{eqnarray}
where terms of ${\cal O}(y^2)$ have been neglected.

By tuning $y$ to be $\ll 1$, we may thus neglect the $y$-corrections to the HC pion 
masses. Note that even if those off-diagonal corrections are numerically 
neglected, 
the HC pions significantly mix independently of the $y$ as in Eq.(\ref{mass-eigenstates:pi}): 
this is the reflection of degenerate perturbation theory well-known 
in the quantum mechanics. 
Such a ``non-decoupling'' mixing will thus affect the HC pion phenomenology as described in 
the next Appendices.

\section{Effective ``Chiral" Lagrangian}
\label{chiral-lag}

In this Appendix we present the effective ``chiral'' Lagrangian 
for the HC pions and derive interaction terms relevant to study 
the LHC phenomenology.

The low-energy effective theory of the present model 
can  be described by the HC pion fields $\Pi$,     
by the nonlinear realization of the underlying flavor ``chiral" 
$SU(3)_L \times SU(3)_R$ symmetry associated with the flavor condensate of $F$-fermions  
$F_{L,R}=(\chi_i, \psi)_{L,R}$ ($i=1,2$), $\langle \bar{\chi}_i \chi_i \rangle = \langle \bar{\psi} \psi \rangle \neq 0$. 
The basic variable to construct the effective model 
is the ``chiral" field $U$, which transforms 
under the global ``chiral" symmetry 
as $U \to g_L \cdot U \cdot g_R^\dag$, 
where $g_{L,R}$ belong to the ``chiral" $SU(3)_{L,R}$ groups, 
respectively. 
When the SM gauges are turned on, 
the global ``chiral" symmetry is partially localized 
according to the SM-gauge embedding as in Ref.~\cite{Haba:2015qbz}. 
Then the effective gauged-``chiral" Lagrangian invariant under the chiral $SU(3)_L \times SU(3)_R$  
and $U(1)_{V}$ symmetries 
is written as  
\begin{eqnarray} 
{\cal L}_{\rm eff} &=& {\cal L}_{\rm kin} + {\cal L}_{\rm mass} + \cdots 
\nonumber \\  
{\cal L}_{\rm kin} &=& 
 \frac{f_\Pi^2}{4} {\rm tr}[|D_\mu U|^2] 
\nonumber \\  
{\cal L}_{\rm mass} &=& b \, {\rm tr}[ U {\cal M}^\dag  +{\rm h.c.}] 
\,, \label{Lag}
\end{eqnarray}
where 
\begin{eqnarray} 
U &=& 
\exp \left( {\frac{2i \Pi}{f_\Pi}} \right) = \exp \left( {\frac{2i \sum_{a=1}^8 \Pi^a \frac{\lambda^a}{2}}{f_\Pi}} \right)  
\,, \nonumber \\ 
 D_\mu U  
&=& 
 \partial_\mu U - i [ {\cal V}_\mu, U ] 
\,, \nonumber \\ 
{\cal V}_\mu &=&  
g_W W_\mu + g_Y B_\mu 
= g_W \sum_{a=1}^3 W_\mu^a \frac{\lambda^a}{2}  
+ g_Y B_\mu Y_F 
\,, \nonumber \\ 
Y_F &=& \frac{\sqrt{3}}{6} \lambda_8 + \frac{1}{\sqrt{6}} \lambda_0 
= \left( 
\begin{array}{ccc} 
1 & & \\ 
& 1 & \\ 
 & & 0
\end{array}
\right) 
\,, \qquad 
\lambda_0 = \frac{2}{\sqrt{6}}{\bf 1}_{3 \times 3} 
\,,  \label{parametrizing}
\end{eqnarray} 
with $\lambda^a$ $(a=1,\cdots ,8)$ being 
the Gell-Mann matrices normalized as ${\rm tr}[\lambda^a \lambda^b]=2 \delta_{ab}$, 
$(W_\mu^a, B_\mu)$ the electroweak gauge fields in the SM, 
and $U$ parametrizes the HC pion fields with regard to the spontaneous breaking  
of the ``chiral'' $SU(3)_L \times SU(3)_R$  symmetry down to the diagonal subgroup $SU(3)_{V}$, just like 
the ordinary QCD, with the associated HC pion decay constant $f_\Pi$. 
The electroweak charges of $U$ have come from the underlying $F$-fermion fields and 
its vector-like condensate~\cite{Haba:2015qbz}. 
In Eq.(\ref{Lag}) two spurion field ${\cal M}$ 
has been 
introduced, in which ${\cal M}$ transforms under the ``chiral" in the same as $U$ 
(${\cal M} \to g_L \cdot {\cal M} \cdot g_R^\dag$). 
The spurion field is assumed to get the vacuum expectation value, 
$\langle {\cal M} \rangle = {\bf 1}_{3 \times 3}$. 
leading to the explicit breaking of the ``chiral" symmetry. 
Then, the ${\cal L}_{\rm mass}$ term can be matched to the underlying 
explicit breaking term, as discussed in the previous section, 
to determine the parameter $b$ in front of it.  
The explicit relation between the parameter $b$ and those explicit breaking coefficients 
will be irrelevant for the present study, so will not be specified here.

In addition to the Lagrangian in Eq.(\ref{Lag}), 
the HC sector yields 
anomalous vertices related to the ``chiral" $SU(3)_L \times SU(3)_R$ 
anomaly with the SM charges gauged, {\it a la} Wess-Zumino-Witten term~\cite{Wess:1971yu}. 
Such terms give significant contributions to HC pion decays to dibosons 
involving photons.  
Taking into account the fact that only vectorial symmetry has been gauged at present, 
one easily finds that only the following term is relevant for the diboson processes:   
\begin{equation} 
 {\cal L}_{\rm WZW} 
 = 
- \frac{N_{\rm HC}}{4 \pi^2 f_\Pi} 
\epsilon^{\mu \nu \rho \sigma} 
{\rm tr}[ 
\partial_\mu {\cal V}_\nu \partial_\rho {\cal V}_\sigma \Pi 
]
\,. 
\end{equation} 
The SM gauge fields in the mass basis $(W^\pm_\mu, Z_\mu, A_\mu)$ 
can be encoded there, by the standard manipulation with Eq.(\ref{Lag}) 
as 
\begin{eqnarray} 
W_\mu^\pm 
&=& 
\frac{W_\mu^1 \mp i W_\mu^2 }{\sqrt{2}} 
\,, \nonumber \\ 
W^3_\mu &=& c_W Z_\mu + s_W A_\mu
\,, \qquad 
B_\mu = - s_W Z_\mu + c_W A_\mu 
\,, 
\nonumber \\ 
s_W &\equiv& \frac{g_Y}{\sqrt{g_W^2 + g_Y^2}} \,, 
\qquad 
c_W^2 \equiv 1 - s_W^2 
\,.        
\end{eqnarray}
in which the electromagnetic coupling $e$ is written 
as $1/e^2 = 1/g_W^2 + 1/g_Y^2$. 
In terms of the mass-eigenstate gauge fields $(W_\mu^\pm, A_\mu, Z_\mu)$, 
the external gauge field ${\cal V}_\mu$ is expressed as 
\begin{eqnarray} 
 {\cal V}_\mu 
 &=& 
\frac{e}{\sqrt{2} s_W} \left( W_\mu^+ I^+ + W_\mu^- I^-  \right) 
\nonumber \\ 
&& 
+ e Q_{\rm em}^F A_\mu 
+ \frac{e}{s_W c_W} \left( I^3 - s_W^2 Q_{\rm em}^F \right) Z_\mu 
 \,, \label{V:mass-eigenstate}
\end{eqnarray}
where 
\begin{eqnarray} 
I^3 &=& 
\frac{\lambda^3}{2} \,, 
\qquad 
I^\pm  = 
\frac{\lambda^1 \pm i \lambda^2}{2} 
\,, \nonumber \\ 
 Q_{\rm em}^F &=& I^3 + Y_F = 
\left( 
\begin{array}{ccc} 
1 & &  \\ 
& 0 &  \\ 
& & 0  
\end{array} 
\right) 
\,,  
\end{eqnarray}

Expanding the $\Pi$ field parametrized as in Eq.(\ref{parametrizing}) in terms of the component fields 
$\Pi^a$ $(a=1,\cdots , 8)$, and using Eq.(\ref{V:mass-eigenstate}), 
one readily finds that the couplings to neutral $\Pi$'s arise as follows:  
\begin{eqnarray} 
 {\cal L}_{\rm WZW}^{\rm NC} &=& 
- \frac{N_{\rm HC}}{4 \pi^2 f_\Pi} 
\sum_{a=3,6,8}
\Bigg[ 
{\rm tr}[I^a \{ I^+, I^-\} ] \cdot \frac{e^2}{2 s_W^2} d W^+ d W^- \Pi^a 
+ 
{\rm tr} [ I^a \{ Q_{\rm em}^F, I^3 - s_W^2 Q_{\rm em}^F  \} ] \cdot \frac{e^2}{s_W c_W} dA dZ \Pi^a
\nonumber \\ 
&& 
+ {\rm tr}[ I^a Q_{\rm em}^F Q_{\rm em}^F] \cdot e^2 d A dA \Pi^a  
- 
{\rm tr}[
I^a  \left(  \{ I^3, Q_{\rm em}^F  \} 
- s_W^2 \cdot Q_{\rm em}^F Q_{\rm em}^F  \right) 
]\ 
\cdot  
\frac{e^2}{c_W^2} d Z d Z  \Pi^a 
\Bigg]  
\nonumber \\
&=& 
- \frac{N_{\rm HC}}{4 \pi^2 f_\Pi} 
\left[ 
\frac{e^2}{2} dA dA + \frac{e^2(c_W^2 - s_W^2)}{2s_W c_W} dA dZ - \frac{e^2}{2} dZdZ \right]   
\left( \Pi^3 + \frac{\Pi^8}{\sqrt{3}} \right) 
\nonumber \\ 
&& 
- \frac{N_{\rm HC}}{4 \pi^2 f_\Pi} 
\left[ 
\frac{e^2}{2s_W^2} dW^+ dW^- \right]   
\left(\frac{\Pi^8}{\sqrt{3}} \right)  
\,, \label{WZW:pi}
\end{eqnarray}
where $dV_1 dV_2 \equiv \epsilon^{\mu\nu\rho\sigma} \partial_\mu V_{1 \nu} \partial_\rho V_{2 \sigma}$, and 
\begin{equation} 
 I^a = \frac{\lambda^a}{2} \,, \qquad {\rm for} \qquad a=1,\cdots , 8  
 \,. 
\end{equation} 
In terms of the mass-eigenstate pions $\{ \tilde{\Pi} \}$ in Eq.(\ref{mass-eigenstates:pi}), 
the WZW interaction terms for the neutral pions 
are expressed as  
\begin{eqnarray} 
 {\cal L}_{\rm WZW}^{\rm NC} 
 &=& - \frac{N_{\rm HC}}{4 \pi^2 f_\Pi} 
 \Bigg[ 
 \left( - \frac{e^2}{2} d A dA + \frac{7e^2}{16s_W^2} d W^+ d W^- - \frac{e^2(c_W^2-s_W^2)}{2s_Wc_W} d A  dZ 
+ \frac{e^2}{2} d Z dZ  \right) \frac{\tilde{\Pi}^3}{\sqrt{3}} 
\nonumber \\ 
&& 
+ \left(  \frac{e^2}{2} d A dA + \frac{3e^2}{16s_W^2} d W^+ d W^- + \frac{e^2(c_W^2-s_W^2)}{2s_Wc_W} d A  dZ 
- \frac{e^2}{2} d Z dZ    \right) \frac{\tilde{\Pi}^6 + \tilde{\Pi}^8}{\sqrt{2}}
 \Bigg] 
 \,. \label{NC:WZW}
\end{eqnarray}
The LHC phenomenology will closely be studied in the next section.

On the other hand, the charged current couplings to the current-eigenstate pions $\{\Pi  \}$ are 
\begin{eqnarray} 
 {\cal L}_{\rm WZW}^{\rm CC} &=& 
- \frac{N_{\rm HC}}{4 \pi^2 f_\Pi} 
\sum_{a=1,2,4,5,7} 
{\rm tr} [
\left(  
\{I^+, Q_{\rm em}^F  \} \, \frac{e^2}{\sqrt{2} s_W} d W^+ d A 
- \{I^+, Q_{\rm em}^F \} \frac{e^2}{\sqrt{2} c_W} d W^+ d Z 
+ {\rm h.c.}  
\right) I^a] \Pi^a 
\nonumber \\ 
&=& 
- \frac{N_{\rm HC}}{4 \pi^2 f_\Pi} 
\left[ 
 \frac{e^2}{2 s_W} d W^+ d A \Pi^- - \frac{e^2}{2 c_W} dW^+ dZ \Pi^- 
 + {\rm h.c.}
\right] 
\,, 
\end{eqnarray}
where 
\begin{equation} 
 \Pi^{\pm} \equiv \frac{\Pi^1 \mp i \Pi^2}{\sqrt{2}} 
 \,. 
\end{equation} 
Writing things in terms of the mass-eigenstates $\{ \tilde{\Pi} \}$ 
with use of Eq.(\ref{mass-eigenstates:pi}), 
one gets the charged-current interaction terms,  
\begin{eqnarray} 
 {\cal L}_{\rm WZW}^{\rm CC} 
&=& 
- \frac{N_{\rm HC}}{4 \pi^2 f_\Pi} 
\left[ 
- \frac{e^2}{2 \sqrt{2} s_W} d W^+ d A (\tilde{\Pi}^- - \tilde{\Pi}^{\prime -} ) + \frac{e^2}{2 \sqrt{2} c_W} dW^+ dZ (\tilde{\Pi}^- - \tilde{\Pi}^{\prime -}) 
 + {\rm h.c.} 
\right] 
\,, \label{CC:WZW}
\end{eqnarray}
where 
\begin{equation} 
 \tilde{\Pi}^{\pm} \equiv \frac{\tilde{\Pi}^1 \mp i \tilde{\Pi}^2}{\sqrt{2}}  
 \,. \qquad 
 \tilde{\Pi}^{\prime \pm} \equiv \frac{\tilde{\Pi}^4 \mp i \tilde{\Pi}^5}{\sqrt{2}} 
\,.   
\end{equation}

In addition to the 8 HC pions, one may write down the WZW term for $\eta'$ coupled to 
the associate current $J_{\mu 5}^0 = \frac{1}{\sqrt{6}} \bar{F} i \gamma_5 F $, in a way similar to $\Pi$'s: 
\begin{eqnarray} 
 {\cal L}_{\rm WZW}^{\eta'} 
 &=& 
- \frac{N_{\rm HC}}{4 \pi^2 f_\Pi} 
\left[ 
e^2  dA dA + \frac{e^2(c_W^2 - s_W^2)}{s_W c_W} dA dZ - e^2  dZdZ \right] \frac{\eta'}{\sqrt{6}} 
\,.    
\end{eqnarray}
Since the $\eta'$ mixes with the pseudoscalar $S$  
through Eq.(\ref{relation:S}),  
in terms of the mass-eigenstates $(s, e_0)$ 
the WZW term for the $\eta'$ now looks like 
\begin{eqnarray} 
 {\cal L}_{\rm WZW}^{\eta'} 
 &\simeq& 
- \frac{N_{\rm HC}}{4 \pi^2 f_\Pi} 
\left[ 
e^2  dA dA + \frac{e^2(c_W^2 - s_W^2)}{s_W c_W} dA dZ - e^2  dZdZ \right] \frac{(g_S s + e_0)}{\sqrt{6}} 
\,,  \label{eta:WZW}  
\end{eqnarray}
 up to terms suppressed by ${\cal O}(g_S^2)$. 
 Here we have omitted the CP-violating terms like $dA dA, dAdZ$ and $dWdW$ since they can be 
 washed out due to the fact that the $SU(2)_W \times U(1)_Y$ groups themselves are topologically trivial.

\section{HC pions at the LHC}
\label{750}
In this Appendix, we shall present quantities relevant for 
the HC pion phenomenologies at the LHC and calculate the HC pion production cross sections. 

\subsection{The decay properties}


 From Eq.(\ref{NC:WZW}) one can easily calculate 
 the partial decay rates for the neutral HC pions $\tilde{\Pi}^{3,6,8}$ to find  
 \begin{eqnarray} 
 \Gamma(\tilde{\Pi}^3 \to \gamma\gamma) 
 &= & 
 \left( \frac{N_{\rm HC} \alpha_{\rm em}}{2 \sqrt{3} \pi f_\Pi} \right)^2 \frac{m_\Pi^3}{16 \pi} 
 \,, \nonumber \\
 \Gamma(\tilde{\Pi}^3 \to WW) 
 &=& 
  \left( \frac{7 N_{\rm HC} \alpha_{\rm em}}{16 \sqrt{3} \pi f_\Pi s_W^2} \right)^2 
\frac{m_\Pi^3}{32 \pi} \left(  1 - \frac{4 m_W^2}{m_\Pi^2} \right)^{3/2} 
 \,, \nonumber \\  
 \Gamma(\tilde{\Pi}^3 \to ZZ) 
 &=& 
  \left( \frac{N_{\rm HC} \alpha_{\rm em}}{2 \sqrt{3} \pi f_\Pi} \right)^2 \frac{m_\Pi^3}{16 \pi} \left(  1 - \frac{4 m_Z^2}{m_\Pi^2} \right)^{3/2} 
 \,, \nonumber \\ 
\Gamma(\tilde{\Pi}^3 \to Z \gamma) 
&=&  
  \left( \frac{N_{\rm HC} \alpha_{\rm em}}{2 \sqrt{3} \pi f_\Pi}\frac{c_W^2 - s_W^2}{s_Wc_W}  \right)^2 
\frac{m_\Pi^3}{32 \pi} \left(  1 - \frac{m_Z^2}{m_\Pi^2} \right)^{3} 
\,, \label{3-widths}
\end{eqnarray} 
and 
\begin{eqnarray}
 \Gamma(\tilde{\Pi}^{6,8} \to \gamma\gamma) 
 &= & 
 \left( \frac{N_{\rm HC} \alpha_{\rm em}}{2 \sqrt{2} \pi f_\Pi} \right)^2 \frac{m_\Pi^3}{16 \pi} 
 \,, \nonumber \\
 \Gamma(\tilde{\Pi}^{6,8} \to WW) 
 &=& 
  \left( \frac{3 N_{\rm HC} \alpha_{\rm em}}{16 \sqrt{2} \pi f_\Pi s_W^2} \right)^2 
\frac{m_\Pi^3}{32 \pi} \left(  1 - \frac{4 m_W^2}{m_\Pi^2} \right)^{3/2} 
 \,, \nonumber \\  
 \Gamma(\tilde{\Pi}^{6,8} \to ZZ) 
 &=& 
  \left( \frac{N_{\rm HC} \alpha_{\rm em}}{2 \sqrt{2} \pi f_\Pi} \right)^2 \frac{m_\Pi^3}{16 \pi} \left(  1 - \frac{4 m_Z^2}{m_\Pi^2} \right)^{3/2} 
 \,, \nonumber \\ 
\Gamma(\tilde{\Pi}^{6,8} \to Z \gamma) 
&=&  
  \left( \frac{N_{\rm HC} \alpha_{\rm em}}{2 \sqrt{2} \pi f_\Pi}\frac{c_W^2 - s_W^2}{s_Wc_W}  \right)^2 
\frac{m_\Pi^3}{32 \pi} \left(  1 - \frac{m_Z^2}{m_\Pi^2} \right)^{3} 
\,, \label{Eq:rate_Pi}
 \end{eqnarray} 
 where $\alpha_{\rm em} \equiv e^2/(4 \pi)$. 
We will hereafter take the mass to be $m_\Pi (= 750 \, {\rm GeV})$ as a reference value. 
Note that the branching fractions of $\tilde{\Pi}^{3,6,8}$ are completely determined 
independently of $N_{\rm HC}$ and $f_\Pi$, 
once the masses and the weak mixing angle are fixed. 
Thus, one gets 
\begin{eqnarray} 
 {\rm Br}(\tilde{\Pi}^3 \to \gamma\gamma)  
&\simeq& 0.10 
\,, \nonumber \\  
{\rm Br}(\tilde{\Pi}^3 \to WW) 
&\simeq& 
0.72 
\,\nonumber \\  
{\rm Br}(\tilde{\Pi}^3 \to ZZ) 
&\simeq& 
0.091
\,, \nonumber \\ 
{\rm Br}(\tilde{\Pi}^3 \to Z\gamma) 
&\simeq& 
0.085
\,,  
\end{eqnarray}
and 
\begin{eqnarray} 
 {\rm Br}(\tilde{\Pi}^{6,8} \to \gamma\gamma)  
&\simeq& 0.24 
\,, \nonumber \\  
{\rm Br}(\tilde{\Pi}^{6,8} \to WW) 
&\simeq& 
0.32
\,\nonumber \\  
{\rm Br}(\tilde{\Pi}^{6,8} \to ZZ) 
&\simeq& 
0.22
\,, \nonumber \\ 
{\rm Br}(\tilde{\Pi}^{6,8} \to Z\gamma) 
&\simeq& 
0.21 
\,.  
\end{eqnarray} 
 The total width is calculated as a function of $N_{\rm HC}$ and $f \equiv f_\Pi/\sqrt{N_{\rm HC}/3}$.  
For $f=$ 92 GeV  
we have 
\begin{eqnarray} 
\begin{array}{c|c|c}
\hspace{20pt} N_{\rm HC} \hspace{20pt} 
& \hspace{20pt} \Gamma_{\rm tot}(\tilde{\Pi}^3)[{\rm MeV}] \hspace{20pt} 
& \hspace{20pt} \Gamma_{\rm tot}(\tilde{\Pi}^{6,8})[{\rm MeV}] \hspace{20pt} \\ 
 \hline  
3  & 46 & 28 \\ 
4  & 61 & 38 \\   
5  & 76 & 47 \\ 
\hline 
\end{array} 
\,.  \label{tot:368}
\end{eqnarray}



The partial decay widths for the 
charged pNG bosons $(\tilde{\Pi}^\pm, \tilde{\Pi}^{\prime \pm})$ are calculated from Eq.(\ref{CC:WZW}) as 
\begin{eqnarray} 
\Gamma(\tilde{\Pi}^{(\prime)\pm} \to W^\pm \gamma) 
&=&  
  \left( \frac{N_{\rm HC} \alpha_{\rm em}}{2 \sqrt{2} \pi f_\Pi s_W} \right)^2 
\frac{m_\Pi^3}{32 \pi} \left(  1 - \frac{m_W^2}{m_\Pi^2} \right)^{3} 
\,, \nonumber \\ 
\Gamma(\tilde{\Pi}^{(\prime)\pm} \to W^\pm Z) 
&=&  
  \left( \frac{N_{\rm HC} \alpha_{\rm em}}{2 \sqrt{2} \pi f_\Pi c_W} \right)^2 
\frac{m_\Pi^3}{32 \pi} \left(  1 - \frac{(m_W +m_Z)^2}{m_\Pi^2} \right)^{3/2} 
\left(  1 - \frac{(m_W - m_Z)^2}{m_\Pi^2} \right)^{3/2} 
\,. 
\end{eqnarray}
Again, the mass has been set to $\simeq 750$ GeV. 
The branching ratios are computed independently of $f_\Pi$ and $N_{\rm HC}$ 
to be 
\begin{eqnarray} 
 {\rm Br}[\tilde{\Pi}^{(\prime) \pm} \to W^\pm \gamma] 
 &\simeq& 0.79 \,, \nonumber \\ 
  {\rm Br}[\tilde{\Pi}^{(\prime)\pm} \to W^\pm Z] 
 &\simeq& 0.21 \,. 
\end{eqnarray}
 For $f=92$ GeV, the total widths are 
\begin{eqnarray} 
\begin{array}{c|c}
\hspace{20pt} N_{\rm HC} \hspace{20pt} 
& \hspace{20pt} \Gamma_{\rm tot}(\tilde{\Pi}^{(\prime) \pm})[{\rm MeV}] \\ 
 \hline  
3  & 19 \\ 
4  & 25 \\   
5  & 32 \\ 
\hline 
\end{array} 
\,.  
\end{eqnarray}


The neutral $\tilde{\Pi}^7$ does not couple in the WZW term as seen from Eq.(\ref{NC:WZW}). 
They may be searched through the multi-body cascade-decay processes like 
$\tilde{\Pi}^7 \to Z^*/\gamma^* + \tilde{\Pi}^{3,6,8} \to l^+l^- + \gamma \gamma $,  
$\tilde{\Pi}^7 \to Z^*/\gamma^* + \tilde{\Pi}^{3,6,8} \to jj + \gamma \gamma $.

\subsection{The LHC Productions and Signals} 
\label{diphoton-cross-HCpions}

The neutral HC pions ($\tilde{\Pi}^{3,6,8}$)  
can dominantly be produced through 
the photon - photon fusion $(\gamma\gamma{\rm F})$ process. 
The 750 GeV resonance production through the $\gamma\gamma$F has been  
studied in Refs.~\cite{Fichet:2015vvy,Csaki:2015vek,Csaki:2016raa,Fichet:2016pvq,Harland-Lang:2016qjy,Barrie:2016ntq,Ahriche:2016mcx,Molinaro:2016oix}. 
We may quote the numerical number estimated in Ref.~\cite{Csaki:2016raa} 
to evaluate  
the $\gamma\gamma$F production of pseudoscalar $\tilde{\Pi}$ 
with the mass $m_{\tilde{\Pi}}=750$ GeV  
at $\sqrt{s}=13(8)$ TeV:  
\begin{equation} 
 \sigma_{\gamma\gamma {\rm F} } 
 (pp \to \tilde{\Pi} \to XY) 
 \simeq   
 10.8 (5.5) \,{\rm pb} \times 
 \left( \frac{\Gamma_{\rm tot}(\tilde{\Pi})}{45 \,{\rm GeV}}  \right) 
 \times {\rm Br}[\tilde{\Pi} \to \gamma\gamma] {\rm Br}[\tilde{\Pi} \to XY] 
 \,,\label{gammagammaF} 
\end{equation} 
where $X$ and $Y$ denote particles produced via the $P$ decays. 
The cross section scales as  
\begin{equation} 
 \sigma_{\gamma\gamma {\rm F}} \propto \frac{N_{\rm HC}^2}{f_\Pi^2} 
\sim \frac{N_{\rm HC}}{f^2} 
\,, \label{scaling}
\end{equation} 
where $f= \frac{f_\Pi}{\sqrt{N_{\rm HC}/3}}$.

Since in the present model all there neutral HC pions $\tilde{\Pi}^{3,6,8}$ contribute 
to the diphoton cross section, the referenced formula in Eq.(\ref{gammagammaF}) 
should be appropriately modified.

First of all, consider the photon-photon scattering amplitudes mediated by $\tilde{\Pi}^{3,6,8}$ 
and write it as $(i {\cal M}_3) + (i {\cal M}_6) + (i {\cal M}_8)$. 
 Taking into account the coupling properties of the neutral HC pions in Eq.(\ref{NC:WZW}), 
we then evaluate the square of the combined scattering amplitude by 
factoring the $\Pi_3$ coupling as   
\begin{equation} 
 \Bigg| (i {\cal M}_3) + (i {\cal M}_6) + (i {\cal M}_8)  \Bigg|^2 
\sim 
\Gamma^2(\tilde{\Pi}^3 \to \gamma\gamma) 
\Bigg| D_3 + 2 \cdot \frac{3}{2} \, D_6  \Bigg|^2 
\,, \label{photon-amp}
\end{equation}
where $D_i = 1/[(M_{\gamma\gamma}^2 - m_\Pi^2) + i m_\Pi \Gamma_{i}]$ with the total 
widths $\Gamma_i$ for $i=3,6,8$ in which $\Gamma_6 = \Gamma_8$ (See Eq.(\ref{tot:368})). 
Using the narrow width approximation, 
\begin{equation} 
|D_i|^2 \approx \frac{\pi}{m_\Pi \Gamma_i} \delta(M_{\gamma\gamma}^2 - m_\Pi^2) 
\,, 
\end{equation}
one can easily rewrite the right hand side of Eq.(\ref{photon-amp}) 
as follows: 
\begin{eqnarray} 
 \Bigg| (i {\cal M}_3) + (i {\cal M}_6) + (i {\cal M}_8)  \Bigg|^2 
&\sim& 
\frac{\pi}{m_\Pi \Gamma_3} \delta(M_{\gamma\gamma}^2 - m_\Pi^2) 
\left[ 1 + \frac{9 \Gamma_3}{\Gamma_6} + \frac{12 \Gamma_3}{\Gamma_6 + \Gamma_3}\right]
\Gamma^2(\tilde{\Pi}^3 \to \gamma\gamma)  
\,. 
\end{eqnarray}
Then, the $\gamma\gamma F$ cross section at the center of mass energy $\sqrt{s}$, 
in which the resonance $\tilde{\Pi}^0$ decays to diphoton, 
is evaluated as  
\begin{eqnarray} 
\sigma_{\gamma\gamma F}^{\gamma\gamma} 
&=& \frac{8 \pi}{s} \int d \eta \int d M_{\gamma\gamma}^2  
\frac{M_{\gamma\gamma}^2}{m_\Pi^2} 
f_{\gamma/p}(\frac{M_{\gamma\gamma}}{\sqrt{s}} e^\eta) 
\cdot 
f_{\gamma/p}(\frac{M_{\gamma\gamma}}{\sqrt{s}} e^{-\eta}) 
\nonumber \\ 
&& \times 
\frac{\pi}{m_\Pi \Gamma_3} \delta(M_{\gamma\gamma}^2 - m_\Pi^2) 
\left[ 1 + \frac{9 \Gamma_3}{\Gamma_6} + \frac{12 \Gamma_3}{\Gamma_6 + \Gamma_3}\right]
\Gamma^2(\tilde{\Pi}^3 \to \gamma\gamma) 
\,,  \label{cross-NWA}
\end{eqnarray}
with the photon luminosity function $f_{\gamma/p}$. 
From the referenced formula in Eq.(\ref{gammagammaF}), 
for $\sqrt{s}=13 (8)$ TeV we read off 
\begin{equation} 
 \frac{8 \pi^2}{s} \frac{1}{m_\Pi} \int d \eta f_{\gamma/p} \cdot f_{\gamma/p} 
= 10.8 (5.5) \,{\rm pb}/(45\,{\rm GeV}) 
\,,  
\end{equation}
so that Eq.(\ref{cross-NWA}) is expressed to be 
\begin{eqnarray} 
\sigma_{\gamma\gamma F}^{\sqrt{s}=13(8)\,{\rm TeV}, \gamma\gamma} 
&=&
10.8 (5.5) \,{\rm pb} 
\times \left( \frac{\Gamma_3}{45\,{\rm GeV}}  \right) 
\times {\rm Br}[\tilde{\Pi}^6 \to \gamma\gamma] {\rm Br}[\tilde{\Pi}^3 \to \gamma\gamma]
\nonumber \\ 
&& 
\times 
\frac{2}{3} 
\left[ 1 + \frac{9 \Gamma_6}{\Gamma_3} + \frac{12 \Gamma_6}{\Gamma_6 + \Gamma_3}\right]
\,, 
\end{eqnarray} 
where we used 
$\Gamma(\tilde{\Pi}^3 \to \gamma\gamma) = 2/3\Gamma( \tilde{\Pi}^6 \to \gamma\gamma)$ 
read off from Eqs.(\ref{3-widths}) and (\ref{Eq:rate_Pi}).

Similarly, one can easily reach the results for the $ZZ$ and $Z \gamma$ channels: 
\begin{eqnarray} 
\sigma_{\gamma\gamma F}^{\sqrt{s}=13(8)\,{\rm TeV}, ZZ/Z \gamma} 
&=&
10.8 (5.5) \,{\rm pb} 
\times \left( \frac{\Gamma_3}{45\,{\rm GeV}}  \right) 
\times {\rm Br}[\tilde{\Pi}^6 \to \gamma\gamma] {\rm Br}[\tilde{\Pi}^3 \to ZZ/Z\gamma]
\nonumber \\ 
&& 
\times 
\frac{2}{3} 
\left[ 1 + \frac{9 \Gamma_6}{\Gamma_3} + \frac{12 \Gamma_6}{\Gamma_6 + \Gamma_3}\right]
\,, 
\end{eqnarray} 
and 
for the $WW$ channel: 
\begin{eqnarray} 
\sigma_{\gamma\gamma F}^{\sqrt{s}=13(8)\,{\rm TeV}, WW} 
&=&
10.8 (5.5) \,{\rm pb} 
\times \left( \frac{\Gamma_3}{45\,{\rm GeV}}  \right) 
\times {\rm Br}[\tilde{\Pi}^6 \to \gamma\gamma] {\rm Br}[\tilde{\Pi}^3 \to WW]
\nonumber \\ 
&& 
\times 
\frac{2}{3} 
\left[ \frac{81}{49} + \frac{\Gamma_6}{\Gamma_3} 
+ \frac{36}{7} \frac{\Gamma_6}{\Gamma_6 + \Gamma_3}\right]
\,, 
\end{eqnarray} 
where use has been made of  
$\Gamma(\tilde{\Pi}^3 \to ZZ/Z\gamma) = 
2/3\Gamma( \tilde{\Pi}^6 \to ZZ/Z \gamma)$ 
and 
$\Gamma(\tilde{\Pi}^3 \to WW) = 
(98/27) \Gamma( \tilde{\Pi}^6 \to WW)$ 
read off from Eqs.(\ref{3-widths}) and (\ref{Eq:rate_Pi}).

Taking $f = 92$ GeV as a reference value, 
below we give lists of the estimated cross sections for the HC pions:

\begin{eqnarray} 
\begin{array}{c|c|c}
\hspace{10pt} N_{\rm HC} \hspace{10pt} 
& \hspace{10pt}  \sigma^{\rm 8TeV}_{\gamma\gamma F}(pp \to \tilde{\Pi}^0 \to \gamma\gamma)[{\rm fb}] \hspace{10pt}  
& \hspace{10pt} \sigma^{\rm 13TeV}_{\gamma\gamma F}(pp \to \tilde{\Pi}^0 \to \gamma\gamma)[{\rm fb}] \hspace{10pt}  
\\ 
 \hline  
3  &  1.3 & 2.5 \\ 
4  & 1.7 & 3.4 \\   
5  & 2.2 & 4.3 \\ 
\hline 
\end{array} 
\,,  \label{tab:pi0-diphoton}
\end{eqnarray}

\begin{eqnarray} 
\begin{array}{c|c|c}
\hspace{10pt} N_{\rm HC} \hspace{10pt} 
& \hspace{10pt}  \sigma^{\rm 8TeV}_{\gamma\gamma F}(pp \to \tilde{\Pi}^0 \to Z\gamma)[{\rm fb}] \hspace{10pt}  
& \hspace{10pt} \sigma^{\rm 13TeV}_{\gamma\gamma F}(pp \to \tilde{\Pi}^0 \to Z\gamma)[{\rm fb}] \hspace{10pt}  
\\ 
 \hline  
3  & 1.1 & 2.2 \\ 
4  & 1.5 & 2.9 \\  
5  & 1.8 & 3.6 \\ 
\hline 
\end{array} 
\,,  
\end{eqnarray}

\begin{eqnarray} 
\begin{array}{c|c|c}
\hspace{10pt} N_{\rm HC} \hspace{10pt} 
& \hspace{10pt}  \sigma^{\rm 8TeV}_{\gamma\gamma F}(pp \to \tilde{\Pi}^0 \to ZZ)[{\rm fb}] \hspace{10pt}  
& \hspace{10pt} \sigma^{\rm 13TeV}_{\gamma\gamma F}(pp \to \tilde{\Pi}^0 \to ZZ)[{\rm fb}] \hspace{10pt}  
\\ 
 \hline  
3  & 1.2 & 2.3 \\ 
4  & 1.6 & 3.1 \\ 
5  & 2.0 & 3.9 \\ 
\hline 
\end{array} 
\,,  
\end{eqnarray}

\begin{eqnarray} 
\begin{array}{c|c|c}
\hspace{10pt} N_{\rm HC} \hspace{20pt} 
& \hspace{10pt}  \sigma^{\rm 8TeV}_{\gamma\gamma F}(pp \to \tilde{\Pi}^0 \to WW)[{\rm fb}] \hspace{10pt}  
& \hspace{10pt} \sigma^{\rm 13TeV}_{\gamma\gamma F}(pp \to \tilde{\Pi}^0 \to WW)[{\rm fb}] \hspace{10pt}  
\\ 
 \hline  
3  & 2.8 & 5.5 \\ 
4  & 3.7 & 7.3 \\   
5  & 4.7 & 9.1 \\ 
\hline 
\end{array} 
\,.   
\end{eqnarray}

The 8 TeV 95\%C.L.limits on 750 GeV scalars decaying to 
$\gamma\gamma$, $Z\gamma$, $WW$ and $ZZ$ have been placed 
as follows~\cite{Aad:2015mna,CMS:2015cwa,Aad:2014fha,Aad:2015kna,Khachatryan:2015cwa,Aad:2015agg}: 
\begin{eqnarray} 
 \sigma_{\gamma\gamma}^{\rm 8\, TeV}|_{\rm exp} 
&\lesssim& 2.3 \,{\rm fb} 
\,, \nonumber \\  
 \sigma_{Z\gamma}^{\rm 8\, TeV}|_{\rm exp}  
&\lesssim& 4.0 \,{\rm fb} 
\,, \nonumber \\  
 \sigma_{ZZ}^{\rm 8\, TeV}|_{\rm exp}  
&\lesssim& 12 \,{\rm fb}
\,, \nonumber \\  
 \sigma_{WW}^{\rm 8\, TeV}|_{\rm exp}  
&\lesssim& 40 \,{\rm fb}  
\,. \label{8TeV:limits}
\end{eqnarray}
Thus, all the predicted signal strengths of $\tilde{\Pi}^{3,6,8}$ 
are consistent with the 8 TeV bounds. 

As to the HC eta-prime, $e_0$, with the mass $={\cal O}(1)$ TeV, 
the prefactors (10.8 and 5.5) in Eq.(\ref{gammagammaF}) are 
changed almost according to the scaling law for the effective photon approximation as 
\begin{equation} 
\frac{ 
\sigma_{\gamma\gamma {\rm F}} (m_{e_0})}{\sigma_{\gamma\gamma {\rm F}}(m_{\tilde{\Pi}}=750\,{\rm GeV})  } 
\approx 
 \left[ \frac{\log(m_{e_0}/\sqrt{s}) }{\log( 750\,{\rm GeV}/\sqrt{s} )} \right]^3 
\simeq 0.73 (0.68)
\,, 
\end{equation} 
with the center of mass energy $\sqrt{s}=13 (8)$ TeV. 
Thus the $e_0$ cross sections are evaluated as 
 \begin{equation} 
 \sigma_{\gamma\gamma {\rm F} } 
 (pp \to e_0 \to XY) 
 \simeq   
 7.8 (3.7) \,{\rm pb} \times 
 \left( \frac{\Gamma_{\rm tot}(e_0)}{45 \,{\rm GeV}}  \right) 
 \times {\rm Br}[e_0 \to \gamma\gamma] {\rm Br}[e_0 \to XY] 
\end{equation}
The partial decay rates are evaluated from Eq.(\ref{eta:WZW}) as  
\begin{eqnarray}
\Gamma \left( e_0 \to \gamma \gamma \right) 
\eqn{=} 
\left( 
\frac{N_{\rm HC} \alpha_{\rm em}}{\sqrt{6} \pi f_\Pi} \right)^2 
\frac{m_{e_0}^3}{16 \pi}\,,\\
\Gamma \left( e_0 \to ZZ \right) 
\eqn{=} 
\left( 
\frac{N_{\rm HC} \alpha_{\rm em}}{\sqrt{6} \pi f_\Pi} \right)^2 
\frac{m_{e_0}^3}{16 \pi} 
\left(
1 - \frac{4 m_Z^2}{m_{e_0}^2} 
\right)^{\frac{3}{2}}\,,\\
\Gamma \left( e_0 \to WW \right) 
\eqn{=} 
\left( 
\frac{N_{\rm HC} \alpha_{\rm em}}{\sqrt{6} \pi f_\Pi s_W^2} \right)^2 
\frac{m_{e_0}^3}{32 \pi} 
\left(
1 - \frac{4 m_W^2}{m_{\eta^0}^2} 
\right)^{\frac{3}{2}}\,,\\
\Gamma \left( e_0 \to Z \gamma \right) 
\eqn{=} 
\left( 
\frac{N_{\rm HC} \alpha_{\rm em}}{\sqrt{6} \pi f_\Pi} \frac{c_W^2 - s_W^2}{s_W c_W} \right)^2 
\frac{m_{e_0}^3}{32 \pi} 
\left(
1 - \frac{m_Z^2}{m_{e_0}^2} 
\right)^{3} \,. 
\end{eqnarray}
and 
\begin{equation} 
\Gamma \left( s \to \gamma \gamma \right) 
= 
\left( 
\frac{g_S N_{\rm HC} \alpha_{\rm em}}{\sqrt{6} \pi f_\Pi} \right)^2 
\frac{m_s^3}{16 \pi} 
\,. 
\end{equation}
The branching ratios for $e_0$ are computed independently of $f_\Pi$ and $N_{\rm HC}$ 
as 
\begin{eqnarray} 
{\rm Br}[e_0 \to \gamma\gamma] 
&\simeq&0.080 
\,, \nonumber \\ 
{\rm Br}[e_0 \to WW] 
&\simeq& 0.77 
\,, \nonumber \\ 
{\rm Br}[e_0 \to ZZ] 
&\simeq& 0.076 
\,, \nonumber \\ 
{\rm Br}[e_0 \to Z \gamma] 
&\simeq& 0.069
\,.  
\end{eqnarray} 
Taking $m_{e_0} = 1$ TeV as a reference value and $f=f_\Pi/\sqrt{N_{\rm HC}/3}=92$ GeV as well, 
we may calculate the total width for $N_{\rm HC}=3,4,5$:   
\begin{eqnarray} 
\begin{array}{c|c}
\hspace{20pt} N_{\rm HC} \hspace{20pt} 
& \hspace{20pt} \Gamma_{\rm tot}(e_0)[{\rm MeV}] \hspace{20pt}  \\ 
 \hline  
3  & 273 \\ 
4  & 364 \\   
5  & 455 \\ 
\hline 
\end{array} 
\,.  
\end{eqnarray}
and the LHC signal strengths of $e_0$ produced via the $\gamma\gamma F$ process:  
\begin{eqnarray} 
\begin{array}{c|c|c}
\hspace{10pt} N_{\rm HC} \hspace{10pt} 
& \hspace{10pt}  \sigma^{\rm 8TeV}_{\gamma\gamma F}(pp \to e_0 \to \gamma\gamma)[{\rm fb}] \hspace{10pt}  
& \hspace{10pt} \sigma^{\rm 13TeV}_{\gamma\gamma F}(pp \to e_0 \to \gamma\gamma)[{\rm fb}] \hspace{10pt}  
\\ 
 \hline  
3  & 0.14 & 0.30 \\ 
4  & 0.19 & 0.40  \\   
5  & 0.24 & 0.51 \\ 
\hline 
\end{array} 
\,.  
\end{eqnarray}
The 8 TeV 95\%C.L.limits on 1 TeV scalars decaying to 
$\gamma\gamma$, $Z\gamma$, $WW$ and $ZZ$ have been placed 
as follows~\cite{Aad:2015mna,CMS:2015cwa,Aad:2014fha,Aad:2015kna,Khachatryan:2015cwa,Aad:2015agg}: 
\begin{eqnarray} 
 \sigma_{\gamma\gamma}^{\rm 8\, TeV}|_{\rm exp} 
&\lesssim& 1.0 \,{\rm fb} 
\,, \nonumber \\  
 \sigma_{Z\gamma}^{\rm 8\, TeV}|_{\rm exp}  
&\lesssim& 1.5 \,{\rm fb} 
\,, \nonumber \\  
 \sigma_{ZZ}^{\rm 8\, TeV}|_{\rm exp}  
&\lesssim& 10 \,{\rm fb}
\,, \nonumber \\  
 \sigma_{WW}^{\rm 8\, TeV}|_{\rm exp}  
&\lesssim& 35 \,{\rm fb}  
\,, \label{8TeV:limits:1TeV}
\end{eqnarray}
which are far above all the predicted signals of the $e_0$, 
to be tested at the LHC 13 TeV in the near future.

With more precise analysis on the $\gamma\gamma$ fusion 
as done in Ref.~\cite{Lebiedowicz:2016lmn}, 
the production cross section can be made 
larger by about factor of 2 than the numbers in Eq.(\ref{gammagammaF}). 
Then, the optimal value of the decay constant $f$ would be made 
larger by about $\sqrt{2}$, i.e., $f\simeq 130$ GeV. 
In that case, we would have the $\Lambda_{\rm HC}\simeq 4 \pi f \simeq 1.6$ TeV, 
where $m_\Pi/\Lambda_{\rm HC} \simeq 0.5$, so the chiral perturbation with respect to 
the HC pion can be more plausible than the present case with $m_\Pi/\Lambda_{\rm HC} \simeq 0.7$.

\end{document}